\documentclass[twocolumn,aps,pra,showpacs,superscriptaddress]{revtex4}
\usepackage[usenames,dvipsnames]{color}

\usepackage{graphicx}
\usepackage{amsmath,amssymb}
\usepackage{color}
\usepackage{graphics}
\usepackage[FIGTOPCAP,raggedright,nooneline]{subfigure}
\usepackage{multirow}
\usepackage{mathtools}
\usepackage{array}
\newcommand{\be}{\begin{equation}}
\newcommand{\ee}{\end{equation}}
\newcommand{\bea}{\begin{eqnarray}}
\newcommand{\eea}{\end{eqnarray}}

\newcommand{\eins}{\openone}

\newcommand{\NN}{\ensuremath{\mathcal{N}}}

\newcommand{\PP}{\ensuremath{\mathcal{P}}}

\newcommand{\ketbra}[1]{\ensuremath{| #1 \rangle \!\langle #1 |}}

\newcommand{\ket}[1]{\ensuremath{|#1\rangle}}

\newcommand{\bra}[1]{\ensuremath{\langle#1|}}

\newcommand{\braket}[1]{\ensuremath{\langle #1\rangle}}
\newcommand{\KetBraO}[3]{\ensuremath{| #1 \rangle_{#3}\langle #2 |}}
\newcommand{\kommentar}[1]{}

\DeclarePairedDelimiter\floor{\lfloor}{\rfloor}

\renewcommand{\vr}{\ensuremath{\varrho}}

\newcommand{\forget}[1]{}

\begin{document}

%%%%%%%%%%%%%%%%%%%%%%%%%%%%%%%%%%%%%%%%%%%%%%%%%%%%%%%%%%%%%%%%%%%

\title{Optimized parameter estimation in the presence of collective phase noise}

%%%%%%%%%%%%%%%%%%%%%%%%%%%%%%%%%%%%%%%%%%%%%%%%%%%%%%%%%%%%%%%%%%%

%%%%%%%%%%%%%%%%%%%%%%%%%%%%%%%%%%%%%%%%%%%%%%%%%%%%%%%%%%%%%%%%%%%
\author{Sanah Altenburg}
\author{Sabine W\"olk}
\affiliation{Naturwissenschaftlich-Technische Fakult\"at,
Universit\"at Siegen,
Walter-Flex-Str.~3,
D-57072 Siegen}
\author{G\'eza T{\'o}th}
\affiliation{Department of Theoretical Physics, University of the Basque Country UPV/EHU,
P.O. Box 644, E-48080 Bilbao, Spain}
\affiliation{IKERBASQUE, Basque Foundation for Science, E-48013 Bilbao, Spain}
\affiliation{Wigner Research Centre for Physics, Hungarian Academy of Sciences, P.O. Box 49,
H-1525 Budapest, Hungary}
\author{Otfried G\"uhne}
\affiliation{Naturwissenschaftlich-Technische Fakult\"at,
Universit\"at Siegen,
Walter-Flex-Str.~3,
D-57072 Siegen}
% \email{otfried.guehne@uibk.ac.at}

%%%%%%%%%%%%%%%%%%%%%%%%%%%%%%%%%%%%%%%%%%%%%%%%%%%%%%%%%%%%%%%%%%%

\date{\today}

%%%%%%%%%%%%%%%%%%%%%%%%%%%%%%%%%%%%%%%%%%%%%%%%%%%%%%%%%%%%%%%%%%%

\begin{abstract}
We investigate phase and frequency estimation with different 
measurement strategies under the effect of collective phase noise. 
First, we consider the standard linear estimation scheme and present 
an experimentally realisable optimization of the initial probe states 
by collective rotations. We identify the optimal rotation angle for 
different measurement times. Second, we show that sub-shot noise 
sensitivity - up to the Heisenberg limit - can be reached in presence 
of collective phase noise by using differential interferometry, where 
one part of the system is used to monitor the noise. For this, not 
only GHZ states but also symmetric Dicke states are suitable.  We 
investigate the optimal splitting for  a general symmetric Dicke 
state at both inputs and discuss possible experimental realisations 
of differential interferometry.

\end{abstract}

%%%%%%%%%%%%%%%%%%%%%%%%%%%%%%%%%%%%%%%%%%%%%%%%%%%%%%%%%%%%%%%%%%%

%, 37.10.Ty}
%03.65.Ta: Foundations of quantum mechanics; measurement theory
%03.65.Ud: Entanglement and quantum nonlocality
%(e.g. EPR paradox, Bell's inequalities, GHZ states, etc.)
%42.50.Xa: Optical tests of quantum theory
%37.10.Ty: ion trapping

%%%%%%%%%%%%%%%%%%%%%%%%%%%%%%%%%%%%%%%%%%%%%%%%%%%%%%%%%%%%%%%%%%%

\maketitle

\section{Introduction}
%%%%%%%%%%%%%%%%%%%%%%%%%%%%%%%%%%%%%%%%%%%%%%%%%%%%%%%%%%%%%%%%%%%
Quantum metrology offers the promise to measure certain parameters 
with a higher precision than using classical resources only. More precisely, given a physical process $\Lambda(\varphi)$ depending on 
a parameter $\varphi$, one can estimate $\varphi$ with higher accuracy, if
the process is applied to an entangled state of $N$ particles instead
of $N$ separate particles in individual states. In the typical case, 
$\varphi$ is a phase acquired by a unitary evolution, which can, using entanglement, be determined with an accuracy of $(\Delta \varphi)^2 \propto 1/N^2$,  the so-called Heisenberg limit (HL). Contrary to that, with separable states the standard quantum limit (SQL) $(\Delta \varphi)^2 \propto 1/N$ is an upper bound on the precision \cite{oldmetrology, Huelga1997, Sorensen2001, Toth2014, Giovannetti2006, GiovanettiScience, GiovanettiPhotonics, Pezze2009}. 

In any real application, however, errors are unavoidable and one has to
ask whether quantum metrology offers an advantage even in the presence
of noise and decoherence. Here, it was realised that
noise can have a detrimental effect \cite{Escher2011, Demkowicz-Dobrzanski2012}. In fact, for generic noise models and estimation 
schemes, where the same unitary evolution is applied to all particles, 
it was shown that the Heisenberg scaling cannot be retained. This does 
not necessarily mean that quantum effects do not offer any advantage 
anymore, but it shows that one has to consider specific situations and
noise models in detail, in order to find the best quantum mechanical
estimation scheme. In fact, it has been shown that for very specific models the Heisenberg scaling can still be achieved \cite{acinmetro} 
and also ideas from quantum error correction can be used to fight 
against noise \cite{ecmetro1, ecmetro2, ecmetro3}. Finally, for 
specific noise models the optimal states for large numbers of particles have been determined \cite{Frowis2014}.

In this paper, we investigate phase and frequency estimation under 
the effect of collective phase noise, which is a typical noise model
for ion trap experiments \cite{Monz2011}. In the first part, we consider 
the standard linear estimation scheme and optimize the initial probe states 
under collective rotations. It turns out that 
even with this optimization the states do not provide a significant advantage over 
separable states, hence new concepts are needed. In the second part, 
we consider differential interferometry (DI) as such an alternative
concept. In DI the time evolution is only applied to a subset of the
particles, while the other particles are used to monitor the noise only. 
This means that the known negative results \cite{Escher2011, Demkowicz-Dobrzanski2012} do not apply. We use the scenario of DI as introduced in Ref.~\cite{Landini2014}, where it was shown already that DI can sometimes be useful for suppressing decoherence. For our noise model, we present a detailed study which states
are optimal and how many particles should be used for applying the
time evolution and how many particles should be used for monitoring
the noise only. It turns out that a Heisenberg scaling can be reached
again. Finally, we briefly discuss possible implementations of DI using trapped ions.

This paper is organized as follows: In Section II  we describe the metrology scheme 
and the noise model that we are using. In Section III
we determine the optimized states for standard interferometry using our 
noise model. Section IV deals with differential interferometry. We explain
the scheme and discuss the optimal states. We also comment on
possible experimental implementations. Finally, we conclude and discuss 
further open problems. In the Appendix we present detailed calculations and derivations.

%%%%%%%%%%%%%%%%%%%%%%%%%%%%%%%%%%%%%%%%%%%%%%%%%%%%%%%%%%%%%%%%%%%%%%%%%%%%%%%%%%%%%%%%%%%%%%%%%%%%
\begin{figure*}
\begin{center}
\subfigure[ ]{\includegraphics[width=0.4\textwidth]{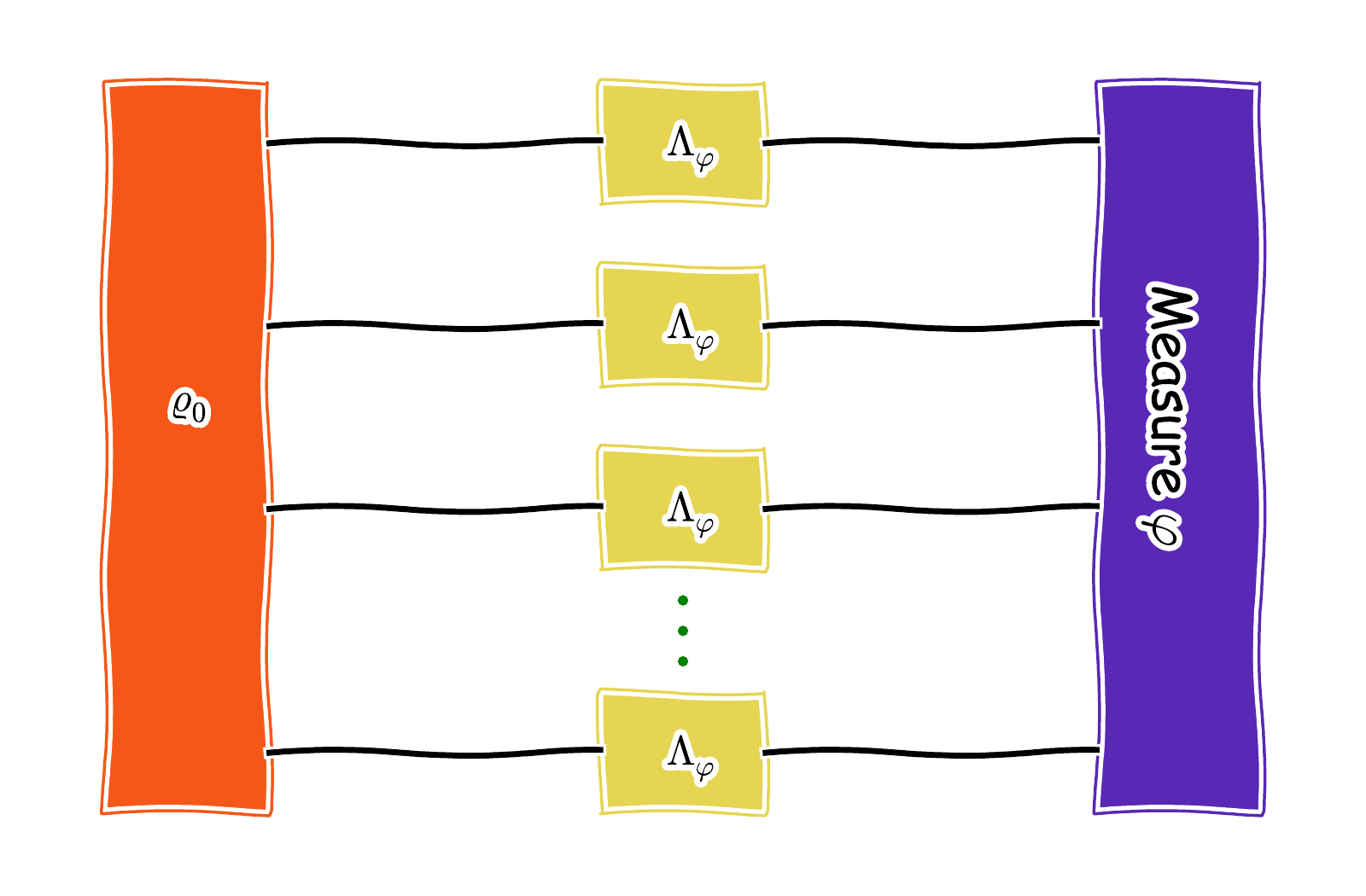}}%\hfill
\subfigure[ ]{\includegraphics[width=0.45\textwidth]{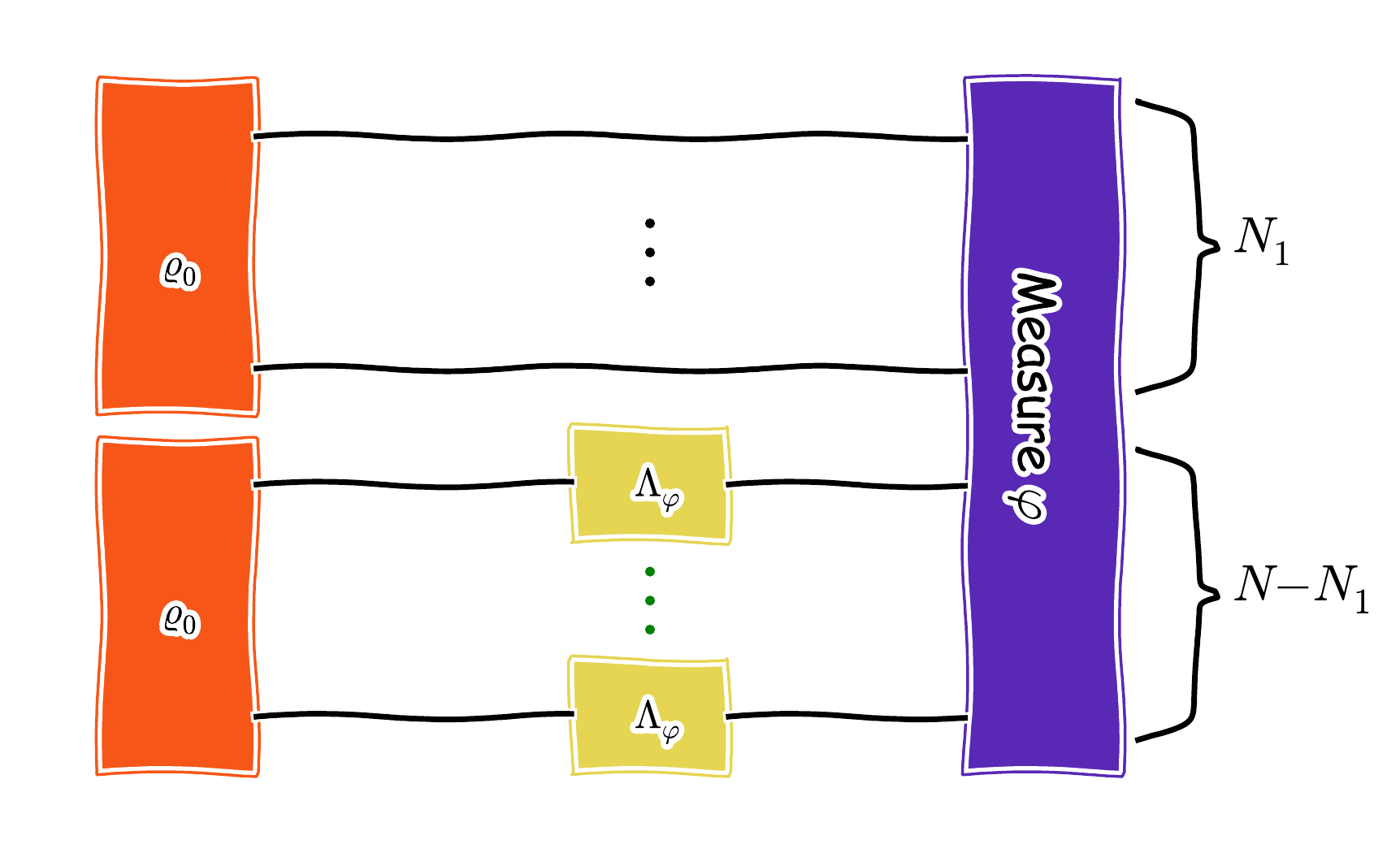}}
\caption{Measurement schemes in quantum metrology.  A map $\Lambda_\varphi$ is acting on each particle individually. The linear map $\Lambda_\varphi$ depends on the parameter $\varphi$. This parameter will be estimated by a measurement. \textbf{(a)}: All particles are initially in a separable or entangled state.  \textbf{(b)}: Differential Interferometry.  The initial state is a bipartite state with $N_1$ particles in the first partition and $N-N_1$ in the second partition.  The linear map $\Lambda_\varphi$ acts on the particles of the second partition only. 
 }\label{fig:Metrology} 
\end{center}
\end{figure*}
%%%%%%%%%%%%%%%%%%%%%%%%%%%%%%%%%%%%%%%%%%%%%%%%%%%%%%%%%%%%%%%%%%%%%%%%%%%%%%%%%%%%%%%%%%

\section{The set-up and the noise model} \label{sec:noise}
In standard metrological schemes (see Fig.~\ref{fig:Metrology}~(a)), 
$N$ particles are in an initial state $\varrho_0$. A time evolution depending on the parameter $\varphi$ acts on each particle individually. The goal is to estimate this parameter $\varphi$ by measurements. 
In classical schemes, the particles are only classically correlated and therefore initially in a separable state. The variance for measuring $\varphi$ is bounded by the so called Standard Quantum Limit (SQL) $(\Delta \varphi)^2 \propto 1/N$. 
In quantum metrology the particles can be entangled.
With such states the Heisenberg Limit (HL) $(\Delta \varphi)^2 \propto 1/N^2$ can be reached theoretically \cite{Sorensen2001, GiovanettiScience, Giovannetti2006, Pezze2009}. 
As a consequence, there is an enhancement in precision by a factor of $1/N$ by using entangled states. 

However, in realistic experiments, noise affects the particles and reduces the entanglement and thereby the enhancement of using entangled states.  
These noise effects arises because the probe system cannot be perfectly separated from their environment. A possible effect is that the energy splitting of the two-level system depends on the noise influenced by the environment. This causes the level splitting to fluctuate in time. An example of such noise effects are magnetic field fluctuations in systems with magnetic field dependent energy splitting.  In the simplest noise model, all qubits receive the same fluctuations, this is also called collective phase noise.

Collective phase noise is, besides micromotion, the main source of noise in experiments with ions as described in Ref. \cite{Monz2011}.
In experiments with atoms, the trapping potential is fluctuating in time. Those fluctuations also cause collective phase noise, which is besides particle loss the main source of noise in experiments with atoms.
 Without loss of generality we assume magnetic field fluctuations in time as noise source in this paper. However, noise due to trapping potential fluctuations can be described  with the same noise model. 
\begin{figure*}
\subfigure[ ]{\hfill \includegraphics[width=0.4\textwidth]{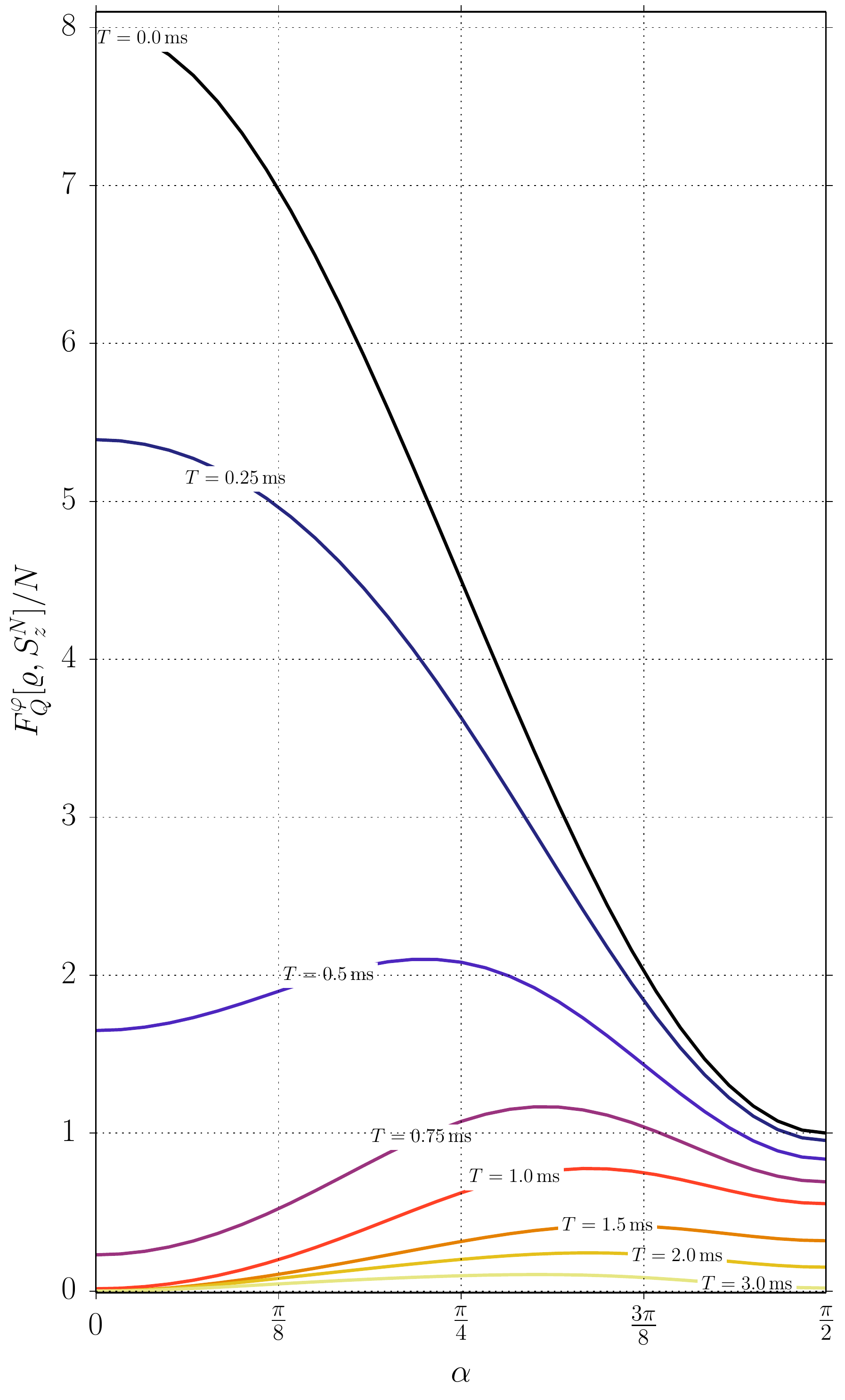}}\hspace{0.5cm}
\subfigure[ ]{\includegraphics[width=0.4\textwidth]{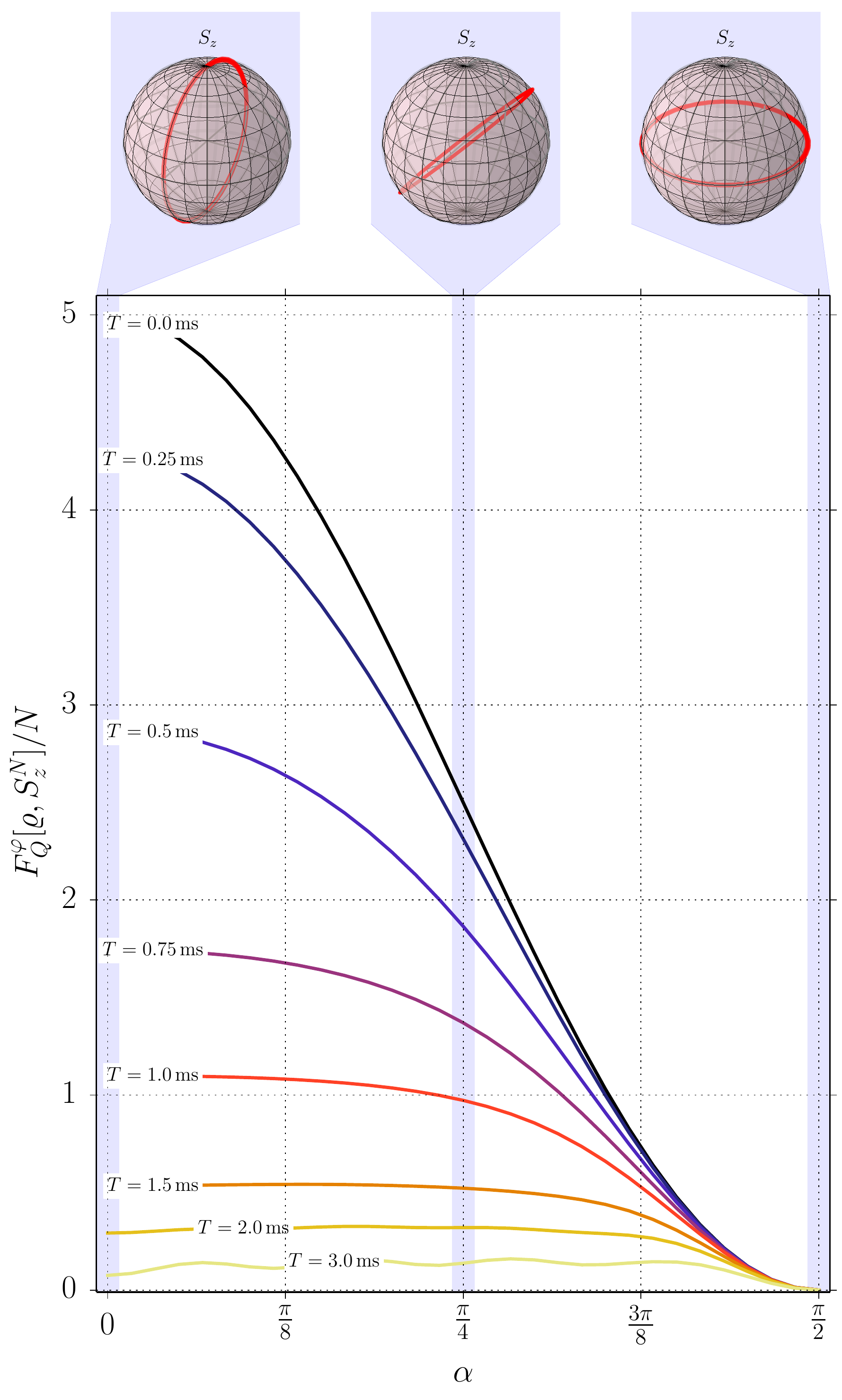}}
\caption{QFI for phase estimation with $N=8$ qubits for different rotated states over rotation angle $\alpha$. Different colors (Color online) represent different measurement times $T$.
\textbf{(a)}: QFI for phase estimation with rotated GHZ states $\ket{\mathrm{GHZ}(\alpha)}$. 
\textbf{(b)}: QFI for phase estimation with rotated symmetric Dicke states $\ket{\mathrm{D}(\alpha)}$. The upper pictures visualize rotated symmetric Dicke states in the Bloch representation.
}\label{fig:Rotation}
\end{figure*}

In a realistic experiment, the Hamiltonian for $N$ particles
with the atomic transition frequency $\omega_0$ and the additional Zeeman splitting due to the magnetic offset field $B_0$ and the magnetic field fluctuations $\Delta B(t)$, is given by
\begin{equation}
H=\hbar (\underbrace{\omega_0+\gamma B_0}_{=\omega}) S_z^N  + \hbar \gamma \Delta B(t) S_z^N,
\end{equation}
with the transition frequency $\omega$.
Here $S_l^N=\sum_{i=1}^N \sigma_l^{(i)}/2$ is the collective spin operator acting on $N$ particles with $l \in \{x,y,z\}$ and the Pauli matrices $\sigma_l^{(i)}$ acting on the $i$-th ion.
The free evolution time $\tau \in \left[0,T\right]$ of the initial state $\vr_0$ can be described by the unitary operator
\begin{equation}
U=\mathrm{exp}\left[-i\left(\omega T + \gamma \int_0^T \mathrm{d}\tau \Delta B(\tau)\right) S_z^N\right].\label{eq:timeevolution}
\end{equation} 
 Here, the magnetic field fluctuations cause phase fluctuations such that the overall phase at a fixed time $T$ is $\Phi = \omega T + \gamma \int_0^T \mathrm{d}\tau \Delta B(\tau)= \omega T + \delta \varphi$. We decompose this unitary into two commuting parts as $U=U_z(\omega T) U_z(\delta\varphi)$, where $U_z(\omega T)$ describes the signal and $U_z(\delta\varphi)$ the noise, with $U_l(\alpha)=\exp\left[-i \alpha S^N_l\right]$.
 The state evolution due to the noise can be described by
 \begin{align}
\bar{\varrho}_T= \braket{U_z(\delta\varphi)\vr_0 U^\dagger_z(\delta\varphi)}_{\delta\varphi} \label{eq:average}
\end{align}
with $\braket{.}_{\delta\varphi}$ denoting the average over all phase fluctuations $\delta\varphi$. 
The final state $\vr$ at a fixed time $T$ is determined by
\begin{align}
\vr&=U_z(\omega T) \braket{U_z(\delta\varphi)\vr_0 U^\dagger_z(\delta\varphi)}_{\delta\varphi} U^\dagger_z(\omega T)\\
& = U_z(\omega T) \bar{\varrho}_T U^\dagger_z(\omega T).\label{eq:average2}
\end{align}
In the following, we make three well justified assumptions, following Ref. \cite{Monz2011}: First of all, we assume  Gaussian phase fluctuations with $\braket{\delta\varphi}_{\delta\varphi}=0$. This means that there is no systematic time dependent bias due to phase fluctuations. Second we assume the time correlation $\braket{\Delta B(t) \Delta B(0)}=\Delta B^2 \exp\left[-t/\tau_c \right] $ to decay exponentially with the correlation time $\tau_c$ and the fluctuation strength $\Delta B$. Third, the noise process can be regarded as stationary $\braket{B(t+\tau)B(t)}=\braket{B(\tau)B(0)}$.

The uncertainty achievable with the help of the time dependent probe state $\varrho(T)$ is lower bounded by the quantum Fisher information (QFI) $F_Q$ via the Cram{\'e}r-Rao bound \cite{Huelga1997, Helstrom1976, Holevo1982,Braunstein1994,Braunstein1996}
\begin{equation}
(\Delta \varphi)^2 \ge \frac{1}{F_Q}.\label{eq:cramer_rao}
\end{equation}
The QFI $F_Q[\varrho_0, \Lambda_\varphi]$ is defined as
\begin{equation}
F_Q[\varrho_0, \Lambda_\varphi]=2\sum_{\alpha,\beta} \frac{|\braket{\alpha|\partial_\varphi \varrho|\beta}|^2}{\lambda_\alpha + \lambda_\beta}\label{eq:QFI}
\end{equation}
with the eigenvalues $\{\lambda_\alpha\}$ and the eigenvectors $\{\ket{\alpha}\}$ of the initial state $\vr_0$.
The QFI does only depend on the initial state and the change of the state $\partial_\varphi \varrho$ due to the linear map $\vr=\Lambda_\varphi (\vr_0)$ and optimizes over all possible measurements.
For the time evolution given by Eq. \eqref{eq:timeevolution}, the QFI for the parameter $\varphi=\omega T$ is given by 
\begin{equation}
F^\varphi_Q[\bar{\varrho}_T,S_z^N]=4\sum_{\alpha<\beta} \frac{(\lambda_\alpha - \lambda_\beta)^2}{\lambda_\alpha + \lambda_\beta}|\braket{\alpha|S_z^N|\beta}|^2
\end{equation}
with the eigenvalues $\{\lambda_i\}$ and the eigenvectors $\{\ket{v_i}\}$ of the averaged state $\bar{\varrho}_T$ given in Eq. \eqref{eq:average}. For the estimation of the frequency $\omega$ we find $F^\omega_Q[\bar{\varrho}_T,T \,S_z^N]=T^2F^\varphi_Q[\bar{\varrho}_T,S_z^N]$.

In the following, we investigate the performance of different probe states depending on time. For this estimate, we assume typical field fluctuations on the order of $\gamma \Delta B=2 \pi\cdot 50\,$Hz and correlation time $\tau_c=1\,$s (see e.g. Ref. \cite{Baumgart2014}).

\section{Phase and frequency estimation with rotated GHZ and symmetric Dicke states} \label{sec:usual_metrology}

In the noiseless case, Greenberger-Horne-Zeilinger (GHZ) states \cite{Greenberger1989} are known to be best for phase estimation in order to reach the HL. Under collective phase noise, they are optimal for frequency estimation, if the measurement time can be optimized \cite{Frowis2014}, which is not always possible. 
They have been realized in several experiments with photons \cite{Bouwmeester1999,Bouwmeester2000} and 
trapped cold ions \cite{Sackett2000,Meyer2001,Monz2011}.
It is known that GHZ states are highly sensitive to particle loss. Losing a particle transforms the state to a separable state, which is useless from a metrological perspective. Dicke states \cite{Dicke1954} are much more robust to particle loss, which makes them interesting for quantum metrology and quantum information processing with BEC's \cite{DickeBEC}, photons \cite{DickePhotons} and trapped cold ions \cite{Schindler2013}.
A simple way to enhance the robustness of GHZ and symmetric Dicke states are collective rotations. Therefore, we will investigate for both, phase and frequency estimation, probe states over collective rotations and test their enhancement in comparison to product states in experiments with collective phase noise.
%%%%%%%%%%%%%%%%%%%%%%%%%%%%%%%%%%%%%%%%%%%%%%%%%%%%%%%%%%%%%%%%%%%%%%%%%%%%%%%%%%%%%%%%%%%%%%%%
%GHZ STATE
%%%%%%%%%%%%%%%%%%%%%%%%%%%%%%%%%%%%%%%%%%%%%%%%%%%%%%%%%%%%%%%%%%%%%%%%%%%%%%%%%%%%%%%%%%%%%%%%
\subsection{GHZ states}
\begin{figure*}
\subfigure[ ]{\includegraphics[width=0.49\textwidth]{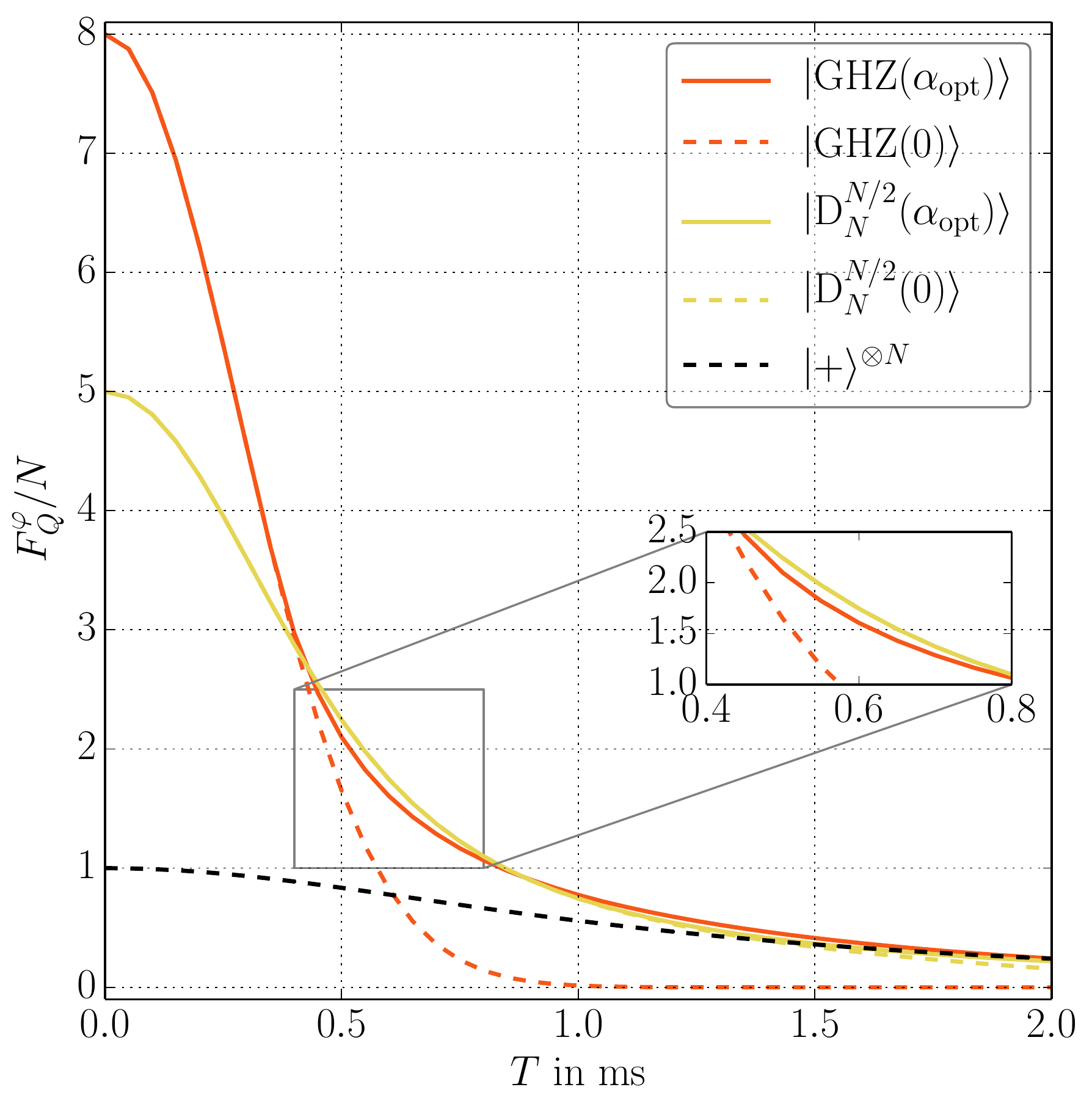}}\hfill
\subfigure[ ]{\includegraphics[width=0.49\textwidth]{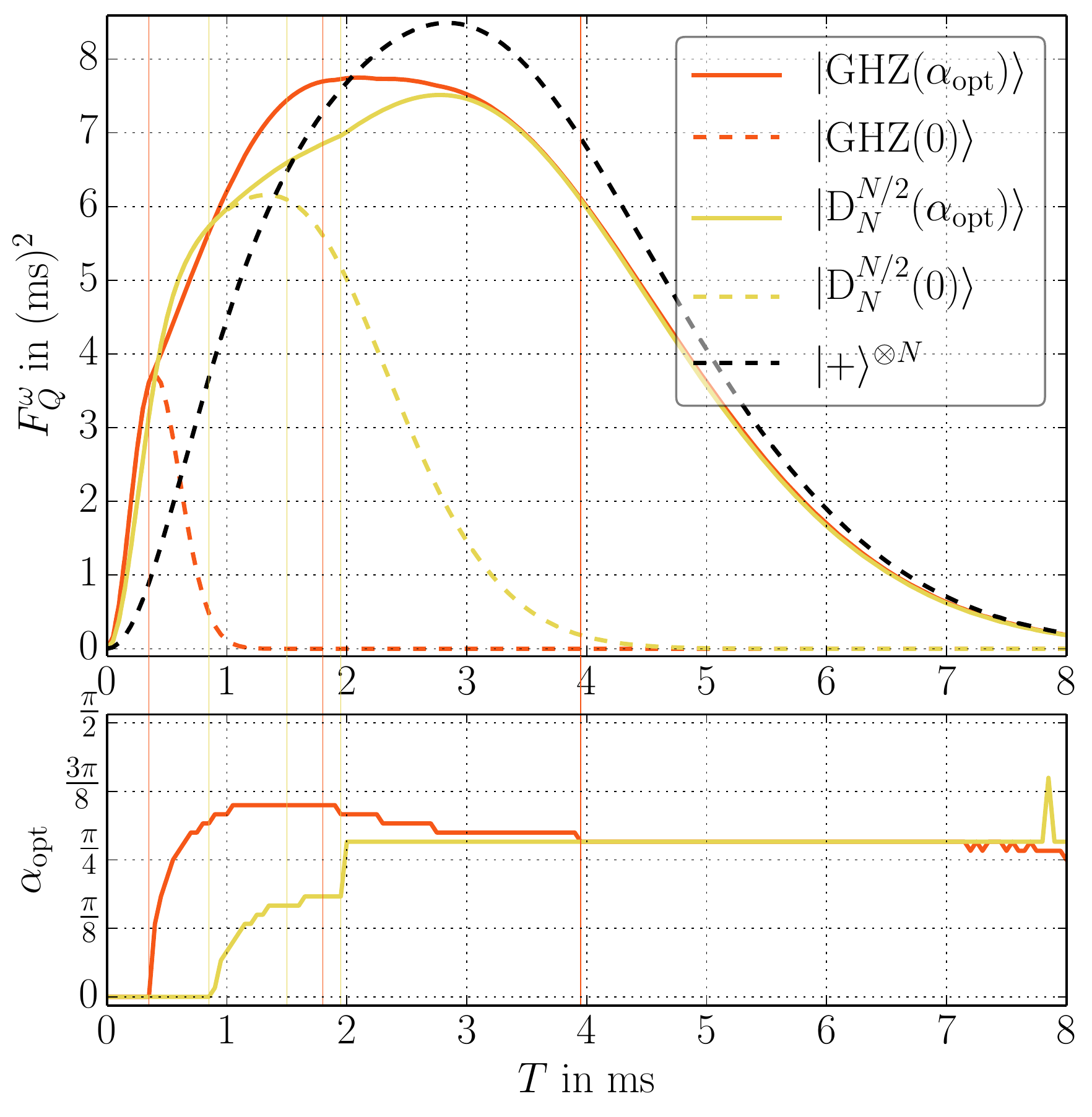}}
\caption{QFI for phase and frequency estimation with $N=8$ qubits.  The solid lines are the QFI optimized over the rotation angle $\alpha$ and dashed lines are the QFI of the origin states.
\textbf{(a):} QFI for phase estimation over the time $T$ for different states.
  \textbf{(b):}  The upper plot shows the  QFI for frequency $\omega$ estimation. The lower plot shows the optimal rotation angle $\alpha_\mathrm{opt}$ over the time for the tested states. 
}\label{fig:Varianceforfreq}
\end{figure*}

The QFI for the GHZ state $\ket{\mathrm{GHZ}}=(\ket{0}^{\otimes N}+\ket{1}^{\otimes N})/\sqrt{2}$ under collective phase noise is given by
\begin{equation}
F_Q^\varphi\left[\bar{\varrho}_T,S_z^N\right]=N^2 \mathrm{e}^{-N^2 C(T)} \label{eq:QFIGHZ}
\end{equation}
with $C(T)=\left(\gamma \Delta B \tau_c\right)^2 \left[\exp(- T/\tau_c)+T/\tau_c -1\right]$ (see Appendix \ref{app:noisy_GHZ} for a detailed calculation). The same result can be obtained by solving the master equations for collective phase noise as has be done in Ref. \cite{Frowis2014} with $(\Delta B)^2=2/(\gamma \tau_c)^2$. This result shows that in the noiseless case, when $T=0$, the HL $F_Q^\varphi = N^2$ can be reached. For $T>0$, the QFI decreases, because the state evolves into a mixed state. The larger $N$, the faster the QFI decreases. For frequency estimation, the QFI increases with $T^2$ for small $T$ and decreases exponentially in time for larger $T$. As a result, there exists an optimal measurement time.

A simple experimentally realizable optimization over the input state are collective rotations
\begin{equation}
 U_y(\alpha)=\mathrm{exp}\left[-i \alpha S_y^N\right]. 
\end{equation} 
 These rotations can be realised with a short laser pulse on all qubits. Due to the symmetry of the state, this rotation can be realised around any axis in the $x/y$-plane. 
Without loss of generality, we choose the $y$-axis, so that the initial state $\varrho_0$ in Eq. \eqref{eq:average} changes to 
\begin{equation}
\varrho_0 \rightarrow U_y(\alpha)\varrho_0 U^\dagger_y(\alpha).
\end{equation}
 We define the rotated GHZ state with $\ket{\mathrm{GHZ}(\alpha)}=U_y(\alpha)\ket{\mathrm{GHZ}}$.
The QFI for phase estimation with $\ket{\mathrm{GHZ}(\alpha)}$ over the rotation angle $\alpha$ is plotted in Fig. \ref{fig:Rotation} (a). It shows the QFI for an $N=8$ GHZ state in comparison to an $N=8$ not rotated product state $\ket{\Psi}$ (dashed lines) for different times $T$. For product states $\ket{\Psi}=\ket{+}^{\otimes N}$ with $\ket{+}=(\ket{0}+\ket{1})/\sqrt{2}$, we find the optimal rotation angle $\alpha_{\mathrm{opt}}=0$ for all $T$. The QFI is symmetric around $\alpha=\pi/2$ because of the symmetry of the state.
For different times $T$ there exists different optimal rotation angles $\alpha_{\mathrm{opt}} \ge 0$ as shown in Fig. \ref{fig:Rotation}(a). The reason is that the state is rotated into a state, which is less sensitive to the magnetic field but also less sensitive to collective phase noise. 

The QFI over time $T$ for the optimal rotation angle $\alpha_{\mathrm{opt}}$ is plotted in Fig. \ref{fig:Varianceforfreq} (a).  Our numerical results show that the QFI for the optimal rotated GHZ state (red solid line) decreases slower than the the not-rotated one (red dashed line) and approaches the QFI for product states (black dashed line) for larger times $T$. 

For frequency estimation there exists a global maximum and a optimal measurement time for all tested states as shown in Fig. \ref{fig:Varianceforfreq}(b). Similar to Ref. \cite{Dorner2012}, we find that product states (dashed black line) perform better then GHZ states (dashed red line) for larger $T$.
However, the measurement time in real experiments is often constrained by external parameters. Therefore, in experiments limited to small measurement times, optimal rotated GHZ states perform better then product states.

%%%%%%%%%%%%%%%%%%%%%%%%%%%%%%%%%%%%%%%%%%%%%%%%%%%%%%%%%%%%%%%%%%%%%%%%%%%%%%%%%%%%%%%%%%%%%%%%
%Dicke state
%%%%%%%%%%%%%%%%%%%%%%%%%%%%%%%%%%%%%%%%%%%%%%%%%%%%%%%%%%%%%%%%%%%%%%%%%%%%%%%%%%%%%%%%%%%%%%%%
\subsection{Symmetric Dicke states}
Symmetric Dicke states with $k$ excitations are defined as
\begin{equation}
\ket{\mathrm{D}^{k}_N}=\frac{1}{\NN} \sum_j \PP_j\{\ket{0}^{\otimes N-k}\otimes\ket{1}^{\otimes k} \},
\end{equation} 
with $\NN$ being a normalization constant and  $\sum_j \PP_j\{.\}$ denoting the sum over all possible permutations.
In experiments with BEC's symmetric Dicke states $\ket{\mathrm{D}^{N/2}_N}$
with $k=\frac{N}{2}$ excitations are often used for quantum metrology, because they are less sensitive to losses (which often appear in such experiments) and still have a good scaling $F_Q\propto N(N+2)/2$ in the noiseless case. 
In the following, we investigate their performance in the presence of collective phase noise.
In general, symmetric Dicke states are insensitive to rotations around the $z$-axis. Therefore, they need to be rotated $\ket{\mathrm{D}}\equiv U_{y}(\pi/2)\ket{\mathrm{D}^{N/2}_N}$, such that the scaling $F\propto N(N+2)/2$ can be achieved in the noiseless case. 
{There are other symmetric Dicke states, which could be metrologically useful as long as $k \propto N$. However, the QFI in the noiseless case is maximal for $k=N/2$. Therefore, we focus on symmetric Dicke states with $k=N/2$ excitations.}

Similar to GHZ states, the state evolves due to collective phase noise, into a mixed state and the QFI decreases in time.
Again, the performance can be enhanced by global rotations $\ket{\mathrm{D}(\alpha)}\equiv U_{y}(\pi/2+\alpha)\ket{\mathrm{D}^{N/2}_N}$. The optimal rotation angles depending on time can be found in Fig. \ref{fig:Rotation} (b). 

The QFI for phase estimation with optimal rotated Dicke states $\ket{\mathrm{D}(\alpha_{\mathrm{opt}})}$ (solid yellow or light grey line) is plotted in Fig. \ref{fig:Varianceforfreq} (a). There is a small enhancement between the QFI for optimal rotated Dicke states $\ket{\mathrm{D}(\alpha_{\mathrm{opt}})}$ and not rotated Dicke states $\ket{\mathrm{D}}$ (dashed yellow or light grey line) for larger $T$.
 We find a small time interval, where optimal rotated Dicke states $\ket{\mathrm{D}(\alpha_{\mathrm{opt}})}$ perform best, that is, also better then optimal rotated GHZ states. 
 For frequency estimation (see Fig. \ref{fig:Varianceforfreq} (b)), not rotated Dicke states  $\ket{\mathrm{D}}$ (dashed yellow or light grey line) perform better than not rotated GHZ states $\ket{\mathrm{GHZ}(0)}$ (dashed red or dark grey line) and product states (black dashed line) perform best. However, there is an enhancement by rotating Dicke states $\ket{\mathrm{D}(\alpha_{\mathrm{opt}})}$ optimal (solid yellow or light grey line).

However, even after optimizing GHZ states and symmetric Dicke states with $N/2$ excitations over rotation angle, product states (black dashed lines) are still the best for frequency estimation if it is possible to tune the measurement time to the optimal one.

In general, the frequency measurement has to be repeated several times and the variance is limited by 
\begin{equation}
 (\Delta\omega)^{-2} \le k F_Q^\omega =t_{o} T F_Q^\varphi
\end{equation}
for $k$ repetitions and the total measurement time $t_o=k T$. If $t_o$ is fixed, GHZ states are optimal for frequency estimation also in presence of collective phase noise, when $T$ can be tuned to it's optimum \cite{Frowis2014}.
In this case, we found the optimal rotated states reach the identical maximum and optimal measurement time $T_\mathrm{opt}$ as the not rotated states and former, after some time $T> T_\mathrm{opt}$ they perform better then the not rotated states. Furthermore, both symmetric Dicke states and GHZ states perform better then product states, when $T$ can be tuned optimal.
However, in experiments with fixed repetition rates $k/t_{o}$, measurement times $T$ are fixed. For such experiments our results in Fig. \ref{fig:Varianceforfreq} become important. From those results, the optimal state at a fixed measurement time can be read out. And we find that there is a time interval, where $\ket{\mathrm{D}(\alpha_{\mathrm{opt}})}$ are optimal, a time interval where $ \ket{\mathrm{GHZ}(\alpha_{\mathrm{opt}})}$ are optimal and for large $T$ product states are optimal.
{This behaviour holds also for large $N$ as shown for $N=50$ in Appendix \ref{app:large_N}.}

In total, we have found, that the GHZ state optimized over the rotation angle has the highest QFI for small times. If it is not possible to measure at small times, another state should be used.  Furthermore, for frequency estimation we find that there is no enhancement in precision by rotating Dicke or GHZ states, if it is possible to measure at the optimal time. However, for smaller measurement times $T$, there is an enhancement by using one of the optimal rotated states. Though, for long measurement times $T$, the QFI for both phase and frequency estimation decreases to zero for all tested states. Therefore, it is important to investigate other metrological schemes.

\section{Differential Interferometry} \label{sec:DI}
In Ref. \cite{Demkowicz-Dobrzanski2012,Escher2011}, it has been shown for a linear interferometer that the enhancement by using entangled states in presence of noise is only a constant factor and not Heisenberg-like. However, 
in Ref. \cite{Landini2014}, it has been shown that with Differential Interferometry (DI) it is possible to reach the HL even in presence of phase noise, the main
mechanism being noise cancellation \cite{Stockton2007}.
DI is a non-linear interferometer for which the results from Ref. \cite{Demkowicz-Dobrzanski2012,Escher2011} do not apply.
DI has been used in many areas of physics, such as measurement of
rotations \cite{Durfee2006}, gradients \cite{Snadden1998} and fundamental
   constants \cite{Fixler2007}.
 So far, DI has been investigated by considering classical Fisher Information with a set of bipartite GHZ states ($\ket{\mathrm{GHZ}}\otimes \ket{\mathrm{GHZ}}$). We will investigate DI for those states by considering QFI and extend this analysis with the class of bipartite symmetric Dicke states.
\begin{figure*}
\subfigure[ ]{\includegraphics[width=0.49\textwidth]{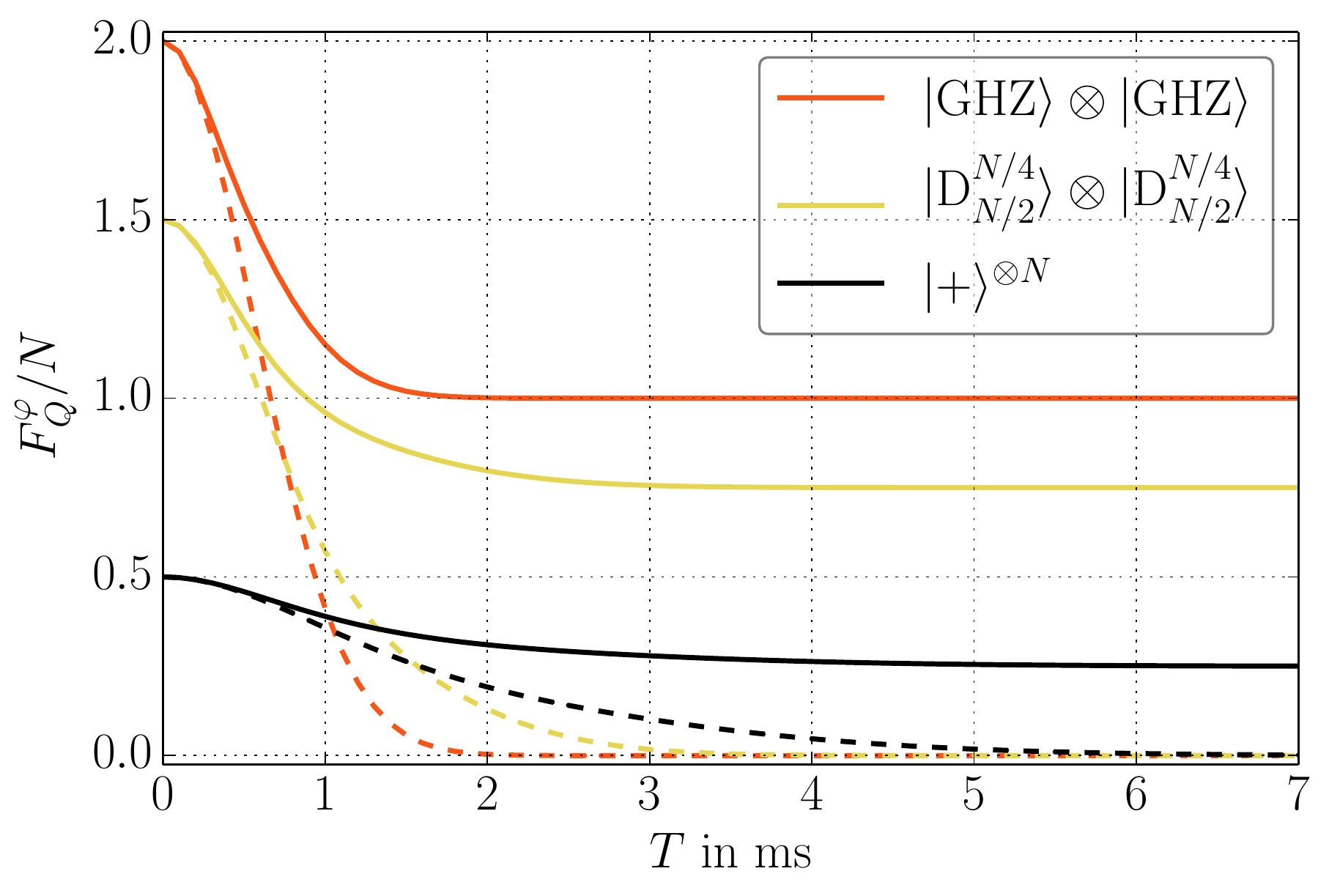}}\hfill
\subfigure[ ]{\includegraphics[width=0.49\textwidth]{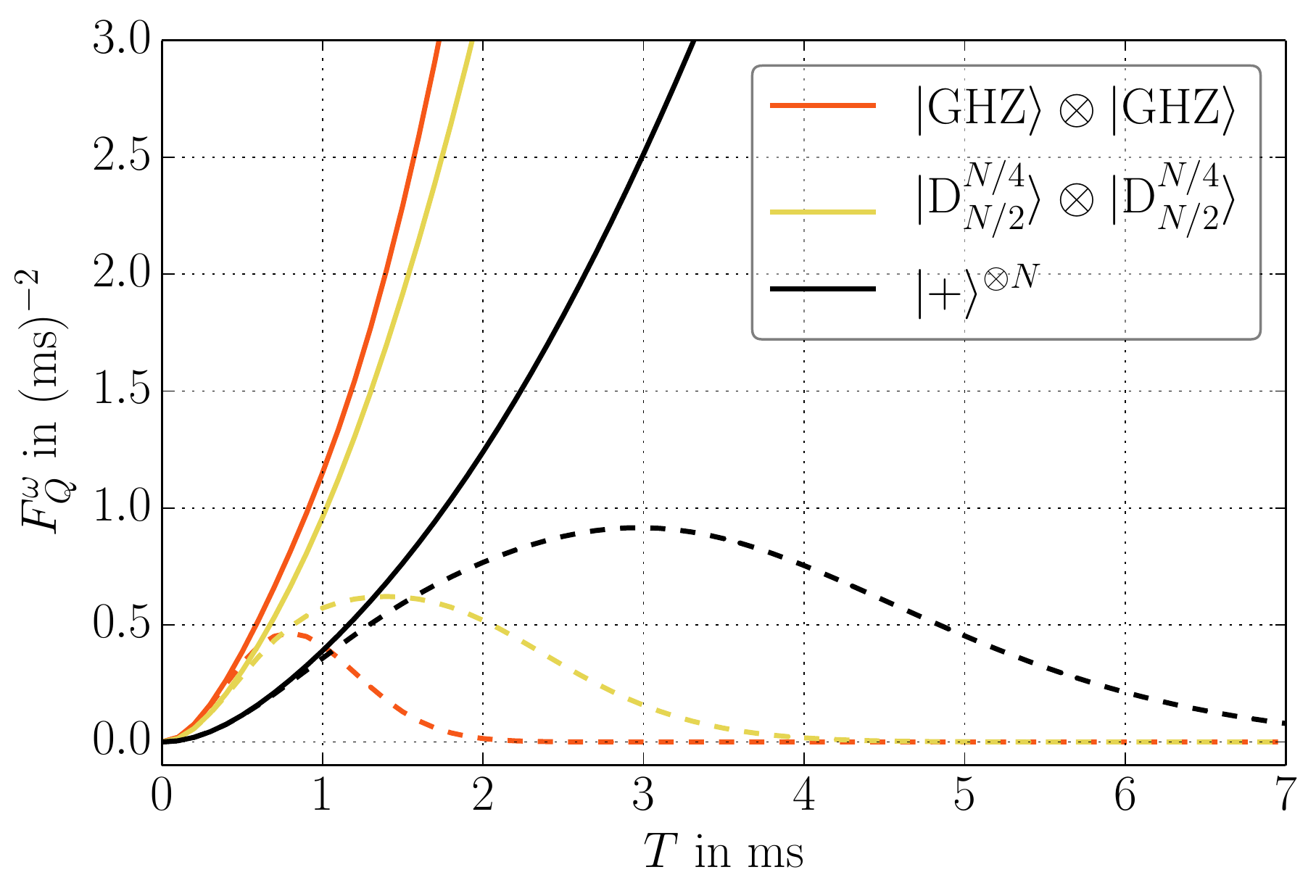}}
\caption{Phase and frequency estimation with equal splitting $N_1=N/2$, by using the ideal DI scheme (solid lines) and DI realised with spin-echo-like experiments (dashed lines), described in Sec. \ref{sec:DI_SE}, with $N=8$ qubits.
\textbf{(a):} QFI for phase estimation over the time $T$ for the tested states.  
  \textbf{(b):} QFI for frequency $\omega$ estimation over the time $T$. 
}\label{fig:phaseDI}
\end{figure*}

In DI, the system is split in two parts.
Both parts will receive the same noise, but only one part will collect the phase $\varphi$ due to a collective rotation around the quantisation axis. This scheme could be interpreted as a measurement of the noise at one part and a measurement of the signal and noise at the other part, such that the noise can be subtracted. It could also be interpreted as a measurement of a phase-difference. The Hamiltonian for this scheme is given by
\begin{equation}
H=\hbar \omega(\eins_{N_1} \otimes S_z^{N-N_1}) + \hbar \gamma \Delta B(t) S_z^N \label{eq:DI}
\end{equation}
with $\eins_{N_1}$ being the identity acting on $N_1$ particles. The last term of Eq. \eqref{eq:DI} describes the noise acting on all particles and the first term is the actual signal. 
In the noiseless case, the maximal QFI is given by \cite{Giovannetti2006}
\begin{equation}
F_Q=4 (\lambda_{\mathrm{max}}-\lambda_{\mathrm{min}})^2=(N-N_1)^2,
\end{equation}
with $\lambda_{\mathrm{max}}$ ($\lambda_{\mathrm{min}}$) being the maximal (minimal) eigenvalue of the generator $\eins_{N_1} \otimes S_z^{N-N_1}$.
This maximal QFI can be reached with the state $ \ket{\Psi}=\left(\ket{v_{\mathrm{max}}}+\ket{v_{\mathrm{min}}}\right)/\sqrt{2}$, where $\ket{v_{\mathrm{max}}}$ and $\ket{v_{\mathrm{min}}}$ are eigenvectors of the generator $\eins_{N_1} \otimes S_z^{N-N_1}$ corresponding to the maximal and respectively minimal eigenvalues.
Optimizing the maximal QFI over the splitting $N_1$ leads to the standard metrological scheme $N_1=0$, discussed in Sec. \ref{sec:usual_metrology}. Here, GHZ states are optimal. However, this state suffers massively from collective phase noise, which leads to $F_Q=0$ for long measurement times leading to the steady state regime.  {Due to noise the state evolves into a mixed state until it becomes a mixture of states from the decoherence free subspace (DFS).
This mixed state does not change due to collective phase noise and is called steady state. }
For the steady state regime the state with maximal QFI is given by (see Appendix \ref{app:DI_optimal_state}) 
\begin{equation}
\ket{\Psi_{\mathrm{opt}}}=\frac{1}{\sqrt{2}}\left(\ket{\underbrace{0  \ldots 0}_{N/2}\underbrace{1  \ldots 1}_{N/2}}+\ket{\underbrace{1  \ldots 1}_{N/2}\underbrace{0  \ldots 0}_{N/2}}\right),\label{eq:DFS}
\end{equation}
with $N_1=N/2$ being optimal.
This state is decoherence free with respect to collective phase noise, such that the QFI for this state is constant in time $F_Q^{\varphi}= N^2/4$ and reaches the HL. 

However, for equal splitting  $N_1=N/2$,  it also has been shown that the state $\ket{\mathrm{GHZ}}\otimes \ket{\mathrm{GHZ}}$ performs good in the presence of correlated phase noise, such that the HL can be reached up to a constant factor. This state contains only $N/2$ particle entanglement, whereas the decoherence free state from Eq. \eqref{eq:DFS} is a genuine multiparticle entangled state. 
In experiments with ions like in Ref. \cite{Monz2011}, the more particle entanglement a state contains the harder the preparation of the state with high fidelity.
 Therefore, we will focus on initial states of the form  
\begin{equation}
\ket{\Psi_N}=\ket{\tilde{\Psi}_{N_1}}\otimes \ket{\tilde{\Psi}_{N-N_1}},\label{eq:DI_initial_states}
\end{equation}
where $\ket{\Psi_N}$ denotes an $N$ particle state, as described in Fig. \ref{fig:Metrology} (b).
We will compare the class of states $\ket{\tilde{\Psi}_{N/2}}=\ket{\mathrm{GHZ}}$ with equal splitting $N_1=N/2$ investigated in \cite{Landini2014} with the class of states given by
\begin{equation}
\ket{\mathrm{D}_{N_1}^{k_1},\mathrm{D}_{N-N_1}^{k_2}}_x=U_y\left(\frac{\pi}{2}\right)\ket{\mathrm{D}_{N_1}^{k_1}} \otimes U_y\left(\frac{\pi}{2}\right)\ket{\mathrm{D}_{N-N_1}^{k_2}},\label{eq:Dicke_state}
\end{equation}
which are bipartite symmetric Dicke (BSD) states in the x basis at both inputs. 

%%%%%%%%%%%%%%%%%%%%%%%%%%%%%%%%%%%%%%%%%%%%%%%%%
%THE OPTIMAL STATE
%%%%%%%%%%%%%%%%%%%%%%%%%%%%%%%%%%%%%%%%%%%%%%%%%%
\subsection{Phase and frequency estimation}
In the following, we will first analyse the scaling behaviour of the here mentioned initial states in DI, that is also the decoherence free case. Then, we will investigate the change of the QFI by adding noise. Finally we will examine the scaling behaviour in the steady state regime.
 
For phase estimation with the initial states and equal splitting $N_1=N/2$, we find % (see Appendix \ref{app:DI_scaling_noiseless}) 
that the QFI scales with $F_Q^\varphi =N/2$ for $\ket{\tilde{\Psi}_{N/2}}$ being product states. For GHZ states $\ket{\tilde{\Psi}_{N/2}}=\ket{\mathrm{GHZ}}$ we find $F_Q^\varphi =N^2/4$ and for the BSD state $\ket{\tilde{\Psi}_{N/2}}=U_{y}(\pi/2)\ket{\mathrm{D}^{N/4}_{N/2}}$ we find $F_Q^\varphi =N(N+4)/8$. 

 In presence of collective phase noise as mentioned in Sec. \ref{sec:noise} the QFI decreases with the time $T$ due to noise as shown in Fig. \ref{fig:phaseDI} (a) for $N=8$.  Nevertheless, the optimal rotation angle $\alpha$ for bipartite GHZ and BSD states is $\alpha_\mathrm{opt}=0$ for all $T$ in DI with equal splitting $N_1=N/2$. 
However, in comparison to the results without DI, for all tested states, the QFI does not decrease to zero. It decreases to a constant value $F_Q^\varphi[\vr_\mathrm{f}]\xrightarrow{} \mathrm{const} > 0$, with $\vr_\mathrm{f}$ being the steady state of the system. 
For frequency estimation we find no maximum for all probe states, such that there is no optimal measurement time. When the QFI for phase estimation becomes constant, that is the steady state regime, the QFI for frequency estimation scales with  $F_Q^\omega \propto T^2$; The larger the measurement time $T$ the better. The QFI for frequency estimation for bipartite GHZ and BSD states, both with $N_1=N/2$, is plotted in Fig. \ref{fig:phaseDI} (b) and we can see that there is an enhancement by using one of the tested entangled states. 

\begin{figure*}
\begin{tabular}{ccc}
\subfigure[ ]{\includegraphics[width=0.3\textwidth]{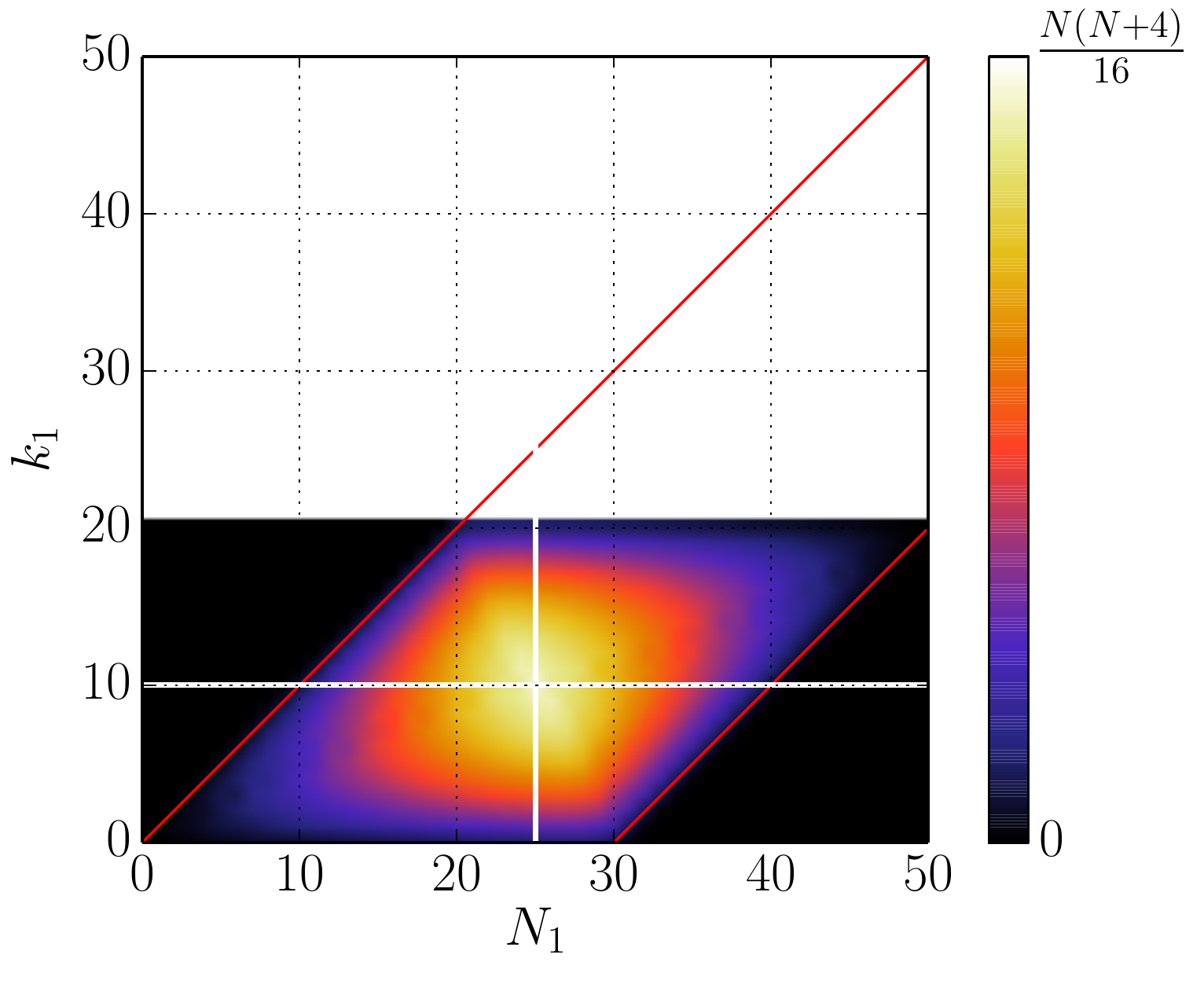}}%\hfill
&
\subfigure[ ]{\includegraphics[width=0.3\textwidth]{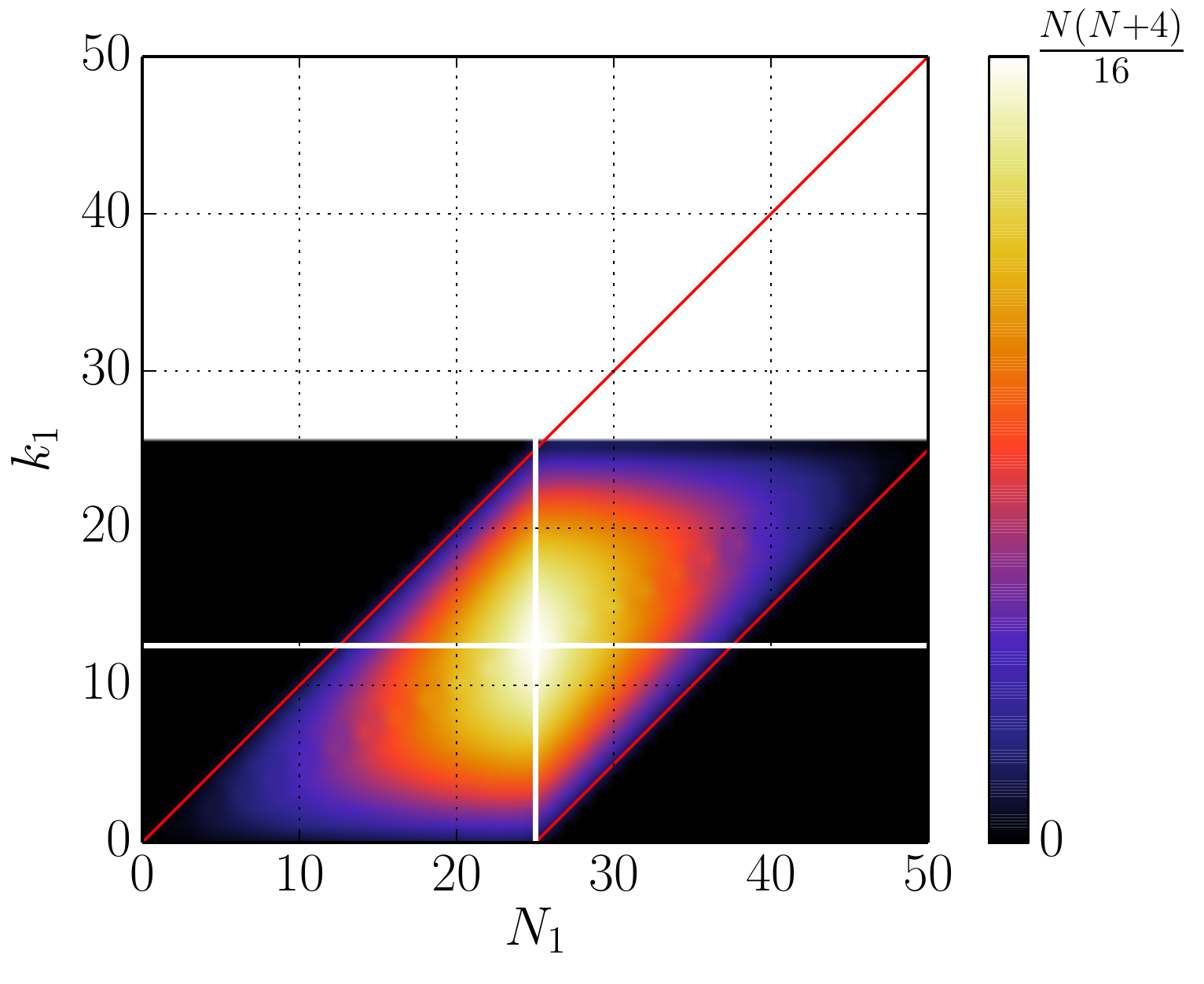}}
&
\multirow{2}{*}{\subfigure[ ]{\includegraphics[width=0.35\textwidth]{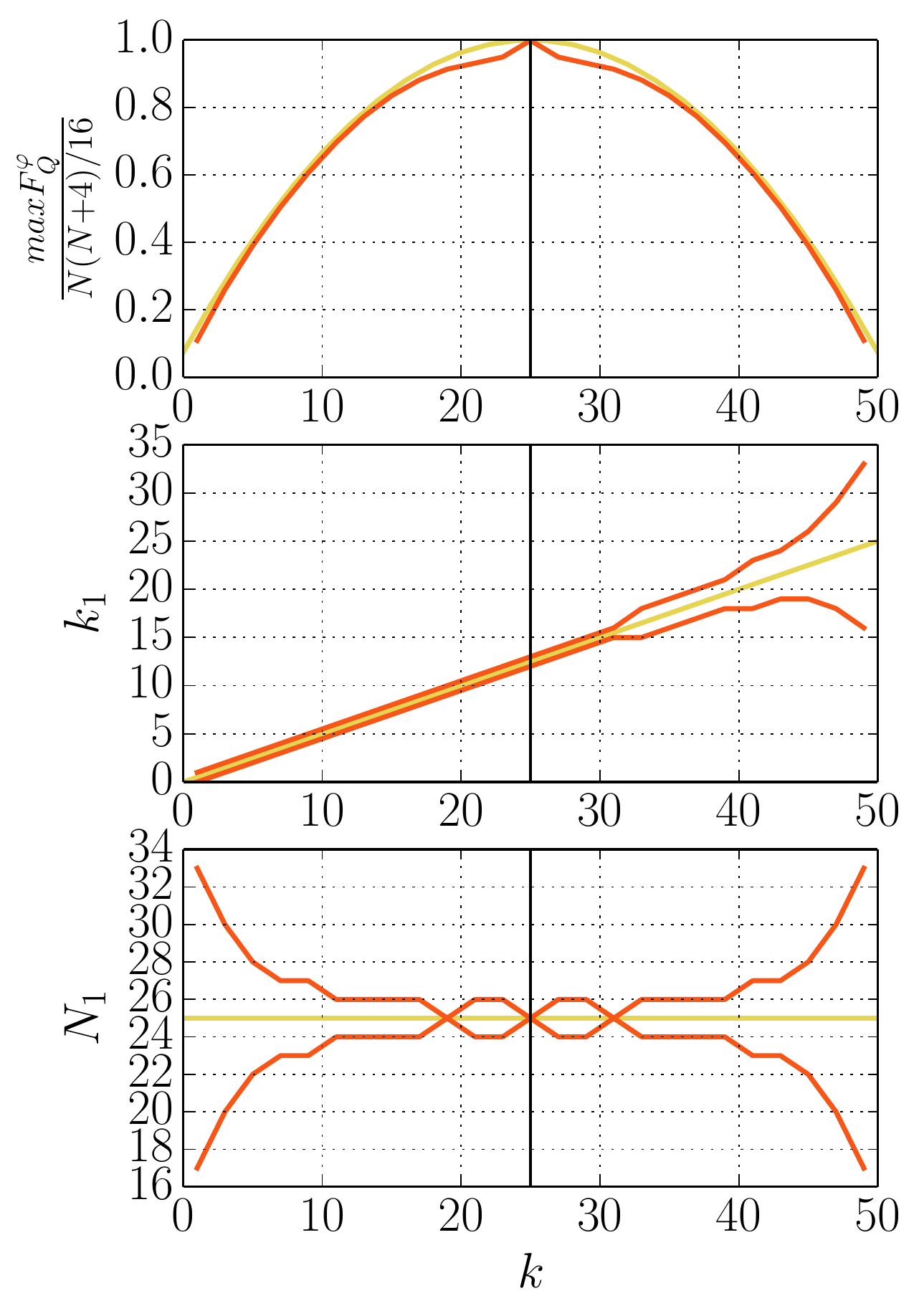}} }
\\
\subfigure[ ]{\includegraphics[width=0.3\textwidth]{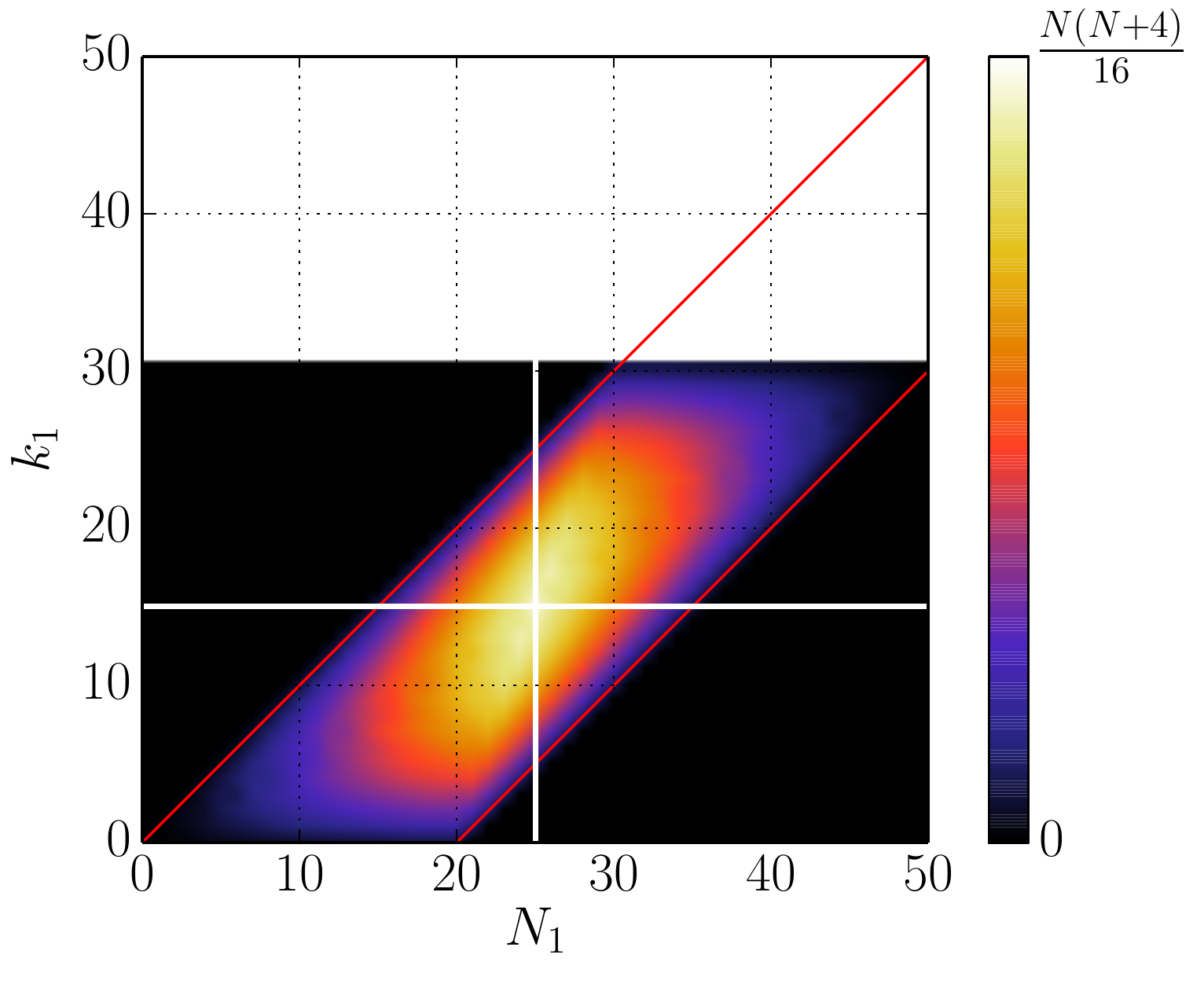}}
&
\subfigure[ ]{\includegraphics[width=0.3\textwidth]{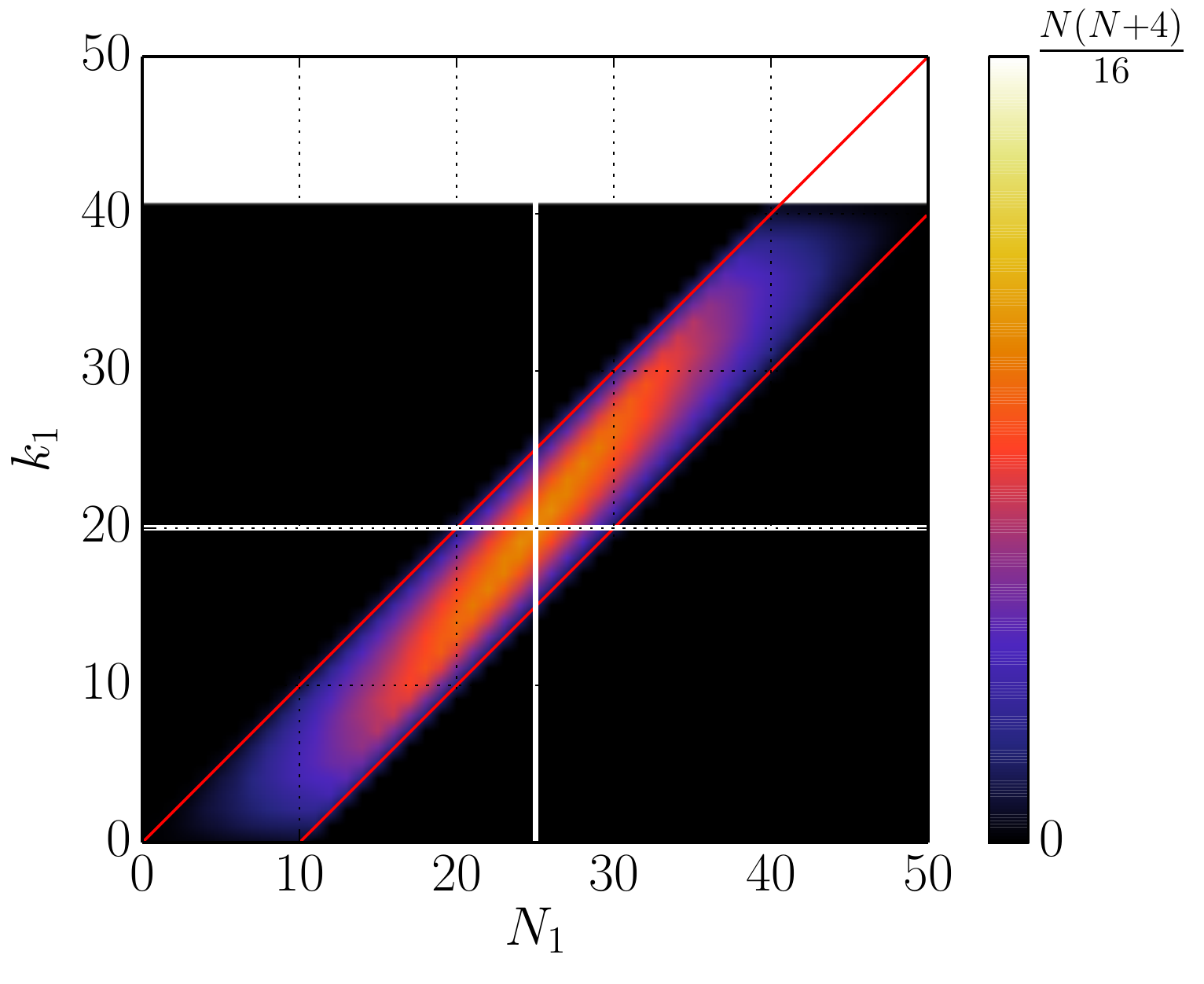}}
\end{tabular}
\caption{QFI (calculated from Eq. \eqref{eq:QFI_for_Dicke1}) of the steady state for an input state $\ket{\mathrm{D}_{N_1}^{k_1}\mathrm{D}_{N-N_1}^{k-k_1}}_x$  by using the metrological scheme described in Fig. \ref{fig:Metrology} (b). Here $N=50$ and is the total number of qubits and $k$ is the total number of excitations. In Fig.  (a), (b), (d) and (e), the QFI as a function of $k_1$ and $N_1$ in shown. In (a) $k=20$, in (b) $k=25$, in (d) $k=30$ and in (e) $k=40$. The white solid lines mark $N/2$ and $k/2$. The red (grey) solid lines margin the area of allowed combinations of $N_1$ and $k_1$ (Color online). In Fig. (c), the maximal QFI is plotted over the total number of excitations $k$, red (dark gray) for odd $k$ and yellow (light gray) for even $k$. The corresponding values for $k_1$ and $N_1$ are shown as a function of $k$. For odd $k$, there are more than one possible combination of $k_1$ and $N_1$ for maximal QFI.}\label{fig:max_FI_Dicke}
\end{figure*}

In the steady state regime, for large $T$, the QFI for phase estimation becomes constant.
 For product states and equal splitting, this constant can be calculated analytically (see Appendix \ref{app:DI_scaling_ss_product}) to
\begin{equation}
F_Q^\varphi[\vr_\mathrm{f}]=N/4.\label{eq:QFI_for_product}
\end{equation}
 For bipartite GHZ states ($\ket{\mathrm{GHZ}}\otimes \ket{\mathrm{GHZ}}$) and equal splitting, this constant can also be calculated analytically (see Appendix \ref{app:DI_scaling_ss_GHZ}) to
 \begin{equation}
 F_Q^\varphi[\vr_\mathrm{f}]=N^2/8.
\end{equation} 
For both, the QFI of the initial state is by a constant factor of two greater than for the steady state. 
For the BSD states, we find (see Appendix \ref{app:DI_scaling_ss_Dicke})
\begin{align}
\begin{split}
F_Q^\varphi[\vr_\mathrm{f}] &= 4 \sum_{k'=0}^{N} \left\lbrace\sum_{q=a}^{b}\left(d^{N_1}_{q,k_1}d^{N-N_1}_{k'-q,k_2}\right)^2\left(k'-q-\frac{N-N_1}{2}\right)^2 \right.  \\
&\left. -\frac{\left[\sum_{q=a}^{b}\left(d^{N_1}_{q,k_1}d^{N-N_1}_{k'-q,k_2}\right)^2 \left(k'-q-\frac{N-N_1}{2}\right)\right]^2}{\sum_{q=a}^{b}\left(d^{N_1}_{q,k_1}d^{N-N_1}_{k'-q,k_2}\right)^2 }  \right\rbrace ,
\end{split}\label{eq:QFI_for_Dicke1}
\end{align}
with $a=\max\{N-N_1-k',0\}$ and $b=\min\{k',N_1\}$.
Here, $d^N_{k',k}\left(\frac{\pi}{2}\right)=\braket{\mathrm{D}_{N}^{k'}|U_y^{N}\left(\frac{\pi}{2}\right)|\mathrm{D}_{N}^{k}}:=d^N_{k',k}$ is the ''small'' Wigner $D$ matrix \cite{Wigner1932} for a rotation angle of $\pi/2$, these are essentially binomial coefficients such, that Eq. \eqref{eq:QFI_for_Dicke1} can directly be evaluated.
For $k_1=k_2=0$ the state in Eq. \eqref{eq:Dicke_state} leads to a product state with splitting $N_1$ and $N-N_1$. We can simplify Eq. \eqref{eq:QFI_for_Dicke1} for that case (see Appendix \ref{app:DI_optimization_product}) and find
\begin{equation}
F_Q^\varphi[\varrho_f]=\frac{N_1(N-N_1)}{N},
\end{equation}
which is maximal for $N_1=\floor{N/2}$ with the maximum $F_Q^\varphi[\varrho_f]=N/4$, which we also found in Eq. \eqref{eq:QFI_for_product}.
For all other possible combinations of $k_1$, $k=k_1+k_2$, $N_1$ and $N-N_1$, we plotted the QFI in Fig. \ref{fig:max_FI_Dicke} for $N=50$. In Fig. \ref{fig:max_FI_Dicke} (c) we plotted the maximal QFI over the total number of excitations $k$, which is proportional to the total energy in the state. For even $k$ (yellow or lighter grey), there is only one maximum for the QFI, whereas for odd $k$ (red or darker gray), there are more than one possible combination of $k_1$ and $N_1$ for maximal QFI. Both, the number of atoms in the first partition $N_1$ and the maximal QFI are symmetric around $k=N/2$. For the number of excitations in the first partition $k_1$ and $k$ being odd, there is no such a symmetry at first sight.
The reason for this asymmetry is the asymmetric splitting for $k<10$ and $k>40$.
However, 
there is a symmetry when comparing the number of excitations in the first partition $k_1$ for $k\le N/2$ with the number of not excited qubits in the first partition $N_1-k_1$ for $k\ge N/2$. Such that $k_1=\floor{k/2}$ is optimal for $k\le N/2$ and $N_1-k_1=\floor{(N-k)/2}$ is optimal for $k\ge N/2$.
The QFI is maximal for $k=\floor{N/2}$, $N_1=\floor{N/2}$ and $k_1=\floor{N/4}$. For $N=4 j$ and $j$ being an integer, this leads to the BSD state $\ket{\mathrm{D}_{N/2}^{N/4},\mathrm{D}_{N/2}^{N/4}}_x$. For that initial state the QFI of the steady state is (see Appendix \ref{app:DI_scaling_ss_Dicke})
 \begin{equation}
 F_Q^\varphi[\vr_\mathrm{f}]=\frac{N(N+4)}{16}.
\end{equation} 
Here again, the QFI of the initial state is by a constant factor of $2$ greater than for the steady state. However, with this steady state, Heisenberg like scaling can be reached. 

We find that bipartite GHZ states with equal splitting are the best for all measurement times $T$. The splitting $N_1=N-N_1=N/2$ is optimal for bipartite GHZ states. If the splitting differs $N_1-N-N_1=N-2N_1 \neq 0$, the steady state is a mixed state, where all coherences vanish, for which $F_Q^\varphi[\vr_\mathrm{f}]=0$.

We find that indeed it is possible to reach the HL with collective phase noise by using DI
and from the tested states, bipartite GHZ states are the best for phase and frequency estimation with this metrological scheme.
We investigated the scaling behaviour for the steady states and found the optimal splitting for bipartite GHZ states to be $N_1=\floor{N/2}$. We also found the optimal probe state out of the set of BSD states that is given by $N_1=\floor{N/2}$ and $k=\floor{N/2}$ and $k_1=\floor{N/4}$. Now, we will discuss possible experimental realisations.

\subsection{Experimental realisation} \label{sec:DI_SE}
An obvious way to realise the operator $\eins_{N/2} \otimes S_z^{N/2}$ seems to be a spin-echo-like experiment on the first $N/2$ particles and a Ramsey-like experiment on the rest of the particles. In a spin-echo-like experiment a $\pi$-pulse flips the spins after the half of the evolution time $T/2$. This flip of the spins induces a rephasing process 
\begin{align}
\begin{split}
&\mathrm{exp}[-i\omega T  S_z^{N/2}- i\gamma \int_0^T \mathrm{d}t \, \omega (t) S_z^{N/2}] \xrightarrow{}\\
& \mathrm{exp}\left[-i\eins_{N/2}-i\gamma \left(\int_0^{T/2}\mathrm{d}t\, \omega (t)-\int_{T/2}^T\mathrm{d}t\, \omega (t)\right) S_z^{N/2} \right]
\end{split}
\end{align}   
with $\omega (t)=\omega+\Delta B(t)$. According to the rephasing process, the signal Hamiltonian changes $H_\mathrm{signal}=\hbar \omega S_z^{N} \xrightarrow{} \hbar \omega (\eins_{N/2}\otimes S_z^{N/2})$. However, the noise on the second part of the particles does not change, but the noise of the first $N/2$ particles changes. It flips its sign after half of the measurement time.
The calculated QFI for phase and frequency estimation is shown in Fig. \ref{fig:phaseDI} (a) and respectively (b). The red or dark gray dashed line shows the behaviour in time $T$, for a bipartite GHZ state, the yellow or light gray dashed line for optimal BSD states $\ket{\mathrm{D}_{N/2}^{N/4},\mathrm{D}_{N/2}^{N/4}}_x$ and the black dashed line for product states. The QFI starts at the same values as with the ideal DI from the previous section. However, it decreases to zero for larger $T$ for both, phase and frequency estimation.
 This means that the advantage of DI gets lost by doing spin-echo-like experiments on one half of the particles. The reason is, that the two parts receive different noise, where as in the ideal DI both parts receive the same noise.
For frequency estimation, we again find an optimal measurement time for all investigated states as shown in Fig. \ref{fig:phaseDI} (b). We also find that the QFI for frequency estimation with all tested probe states decreases to zero after reaching it's maximum. 
Furthermore, for all measurement times, there is no enhancement by the presented scheme in comparison to the usual metrologic scheme, discussed in section \ref{sec:usual_metrology}.
This means that the presented scheme is insufficient for realising DI.

Another idea, which will turn out to be wrong, to realise the metrological scheme from Fig. \ref{fig:Metrology} (b) would be to repeat the experiment: One time with noise and signal and one time with noise only.
For this method the desired signal Hamiltonian can be realised, but the averaged state changes
\begin{equation}
\braket{.\otimes .}_{\delta\varphi} \xrightarrow{} \braket{.}_{\delta\varphi} \otimes \braket{.}_{\delta\varphi}.
\end{equation}
In this case, the steady state has no coherences, with respect to the bipartition of the Hamiltonian, left. This means, that for all probe states, the QFI for phase estimation will decrease to zero. Therefore, this method is also insufficient for realising DI.

%%%%%%%%%%%%%%%%%%%%%%%%%%%%%%%%%%%%%%%%%%%%%%%%%%%%%%%%%%%
\section{Conclusions}
%%%%%%%%%%%%%%%%%%%%%%%%%%%%%%%%%%%%%%%%%%%%%%%%%%%%%%%%%%%

We investigated the usual metrological 
scheme and differential interferometry with a set of prominent 
probe states in presence of collective phase noise. For standard
metrology schemes we determined the optimized states. Then we
showed that with differential interferometry it is possible to reach a good 
scaling - up to the  Heisenberg limit - even in presence of collective 
phase noise. Here, from the tested set of bipartite probe states, 
bipartite GHZ states are optimal for both phase and frequency estimation.
However, GHZ states are highly sensitive to particle losses. Therefore, 
in experiments where particle losses appear frequently, symmetric 
Dicke states are often used. We found that with bipartite symmetric
Dicke states it is also possible to reach a good scaling up to  
Heisenberg scaling.

As we have seen, however, differential interferometry may be hard to 
realise in experiments. Therefore, it would be useful to design 
experimentally feasible schemes for implementing these ideas. In addition, an extension of the differential method to 
other metrology schemes, e.g., the measurement of oscillating fields
\cite{Baumgart2014} is highly desirable.

\vspace{0.2cm}
\noindent
We thank 
I. Appelaniz,
M. Johanning,
J. Kolodinski, 
M. Mitchell,
M. Oszmaniec,
L. Pezz\`e, 
A. Smerzi, 
P. Treutlein,
G. Vitagliano, 
and
Ch. Wunderlich 
for discussions. 
This work has been supported by the Friedrich-Ebert-Stiftung, 
the ERC (Consolidator Grant 683107/Tempo and Starting Grant
258647/GEDENTQOPT), the EU (CHIST-ERA QUASAR, COST Action 
CA15220), the MINECO (Projects Nos. FIS2012-36673-C03-03
and FIS2015-67161-P), the Basque Government (Project
No. IT4720-10), the OTKA (Contract No. K83858),
the UPV/EHU program UFI 11/55, the FQXi Fund
(Silicon Valley Community Foundation), and the DFG.

\onecolumngrid

\begin{appendix}
\section{GHZ state under collective phase noise}\label{app:noisy_GHZ}
As described in section \ref{sec:noise}, the $N$ particle GHZ state evolves, due to the collective phase noise, at a certain time $t$ into the (over phase fluctuations) averaged state
\begin{align}
\bar{\varrho}(t)=\frac{1}{4} \ket{0^{\otimes N}}\bra{0^{\otimes N}}+\frac{1}{4} \ket{1^{\otimes N}}\bra{1^{\otimes N}}+ \frac{d(t)}{4} \ket{0^{\otimes N}}\bra{1^{\otimes N}}+ \frac{d(t)}{4}\ket{1^{\otimes N}}\bra{0^{\otimes N}}
\end{align}
with $d(t)=\mathrm{exp}\left[-\frac{1}{2}\left(N \gamma \Delta B \tau_c\right)^2 \left(\exp(- t/\tau_c)+t/\tau_c -1\right)\right]$.  The mixed state has non-zero eigenvalues $\lambda_{\pm}=\frac{1 \pm d(t)}{2}$ and corresponding eigenvectors $\ket{\pm}=\frac{1}{\sqrt{2}} (\ket{0^{\otimes N}} \pm \ket{1^{\otimes N}})$. We denote all other eigenvalues with $\lambda_i=0$ and the corresponding eigenvectors $\ket{v_i}$ such that we can rewrite the state as
\begin{equation}
\bar{\varrho}(t)= \frac{1 + d(t)}{2}\ketbra{+}+\frac{1 - d(t)}{2}\ketbra{-}.
\end{equation}
With that state, we can calculate the QFI for phase estimation by using the metrological scheme from Fig. \ref{fig:Metrology}.
Therefore, we use the fact that $S_z^N\ket{+}=\frac{N}{2}\ket{-}$ such that $\braket{v_i|S_z^N|\pm}=0$ to calculate the QFI and arrive at
\begin{align}
\begin{split}
F^\varphi_Q&=4\sum_{\alpha<\beta} \frac{(\lambda_\alpha - \lambda_\beta)^2}{\lambda_\alpha + \lambda_\beta}|\braket{\alpha|S_z^N|\beta}|^2\\
&=4\frac{(\lambda_+ - \lambda_-)^2}{\lambda_+ + \lambda_-}|\braket{+|S_z^N|-}|^2 =N^2 d(t)^2.
\end{split}
\end{align}
For frequency estimation we find
\begin{equation}
F^\omega_Q=t^2N^2 d(t)^2.
\end{equation}
These results for the QFI are similar to the ones in Ref. \cite{Frowis2014}.

\section{Optimal rotation angle for $N=50$ qubits}\label{app:large_N}
\begin{figure*}
\subfigure[ ]{\includegraphics[width=0.49\textwidth]{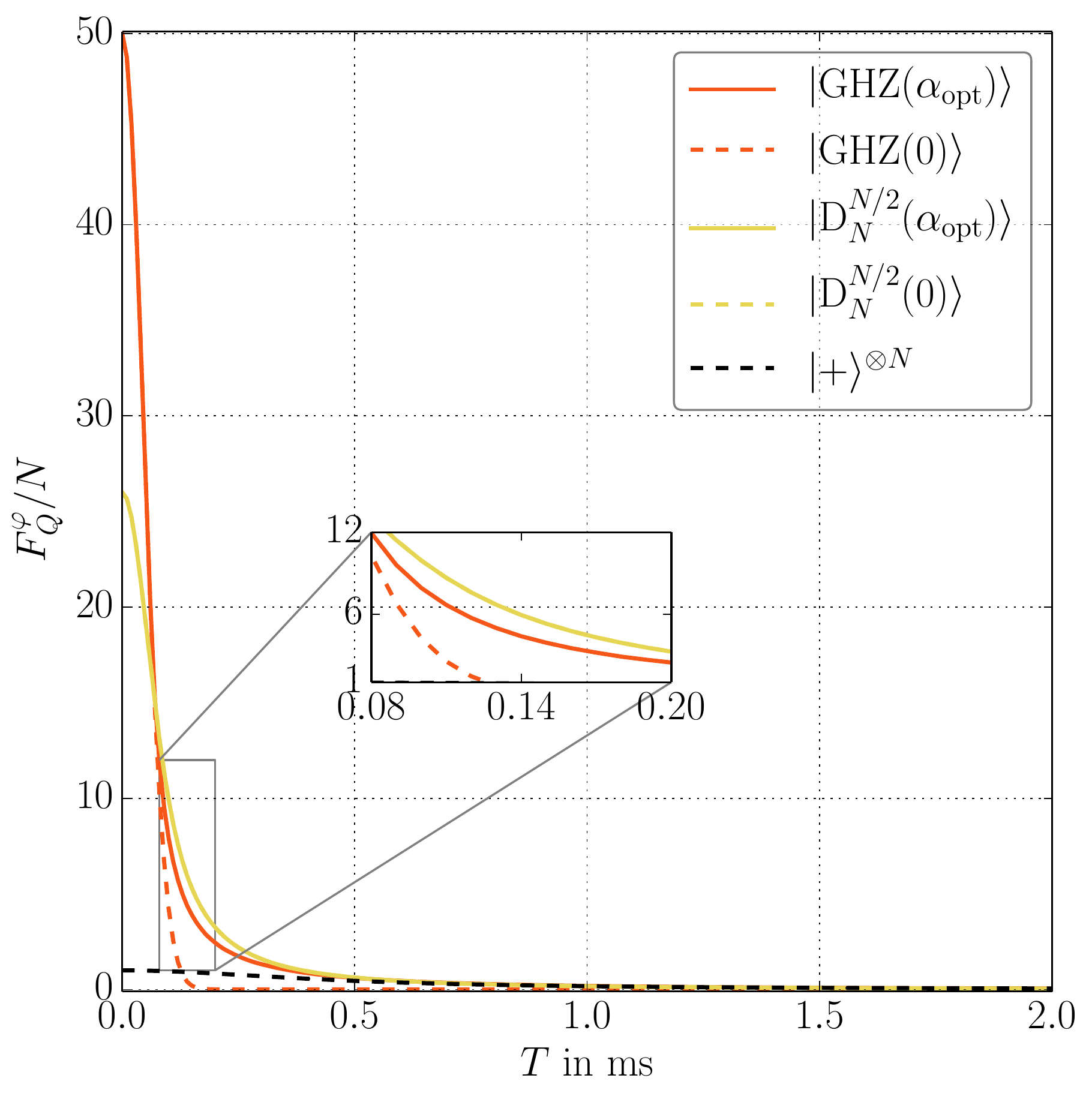}}\hfill
\subfigure[ ]{\includegraphics[width=0.49\textwidth]{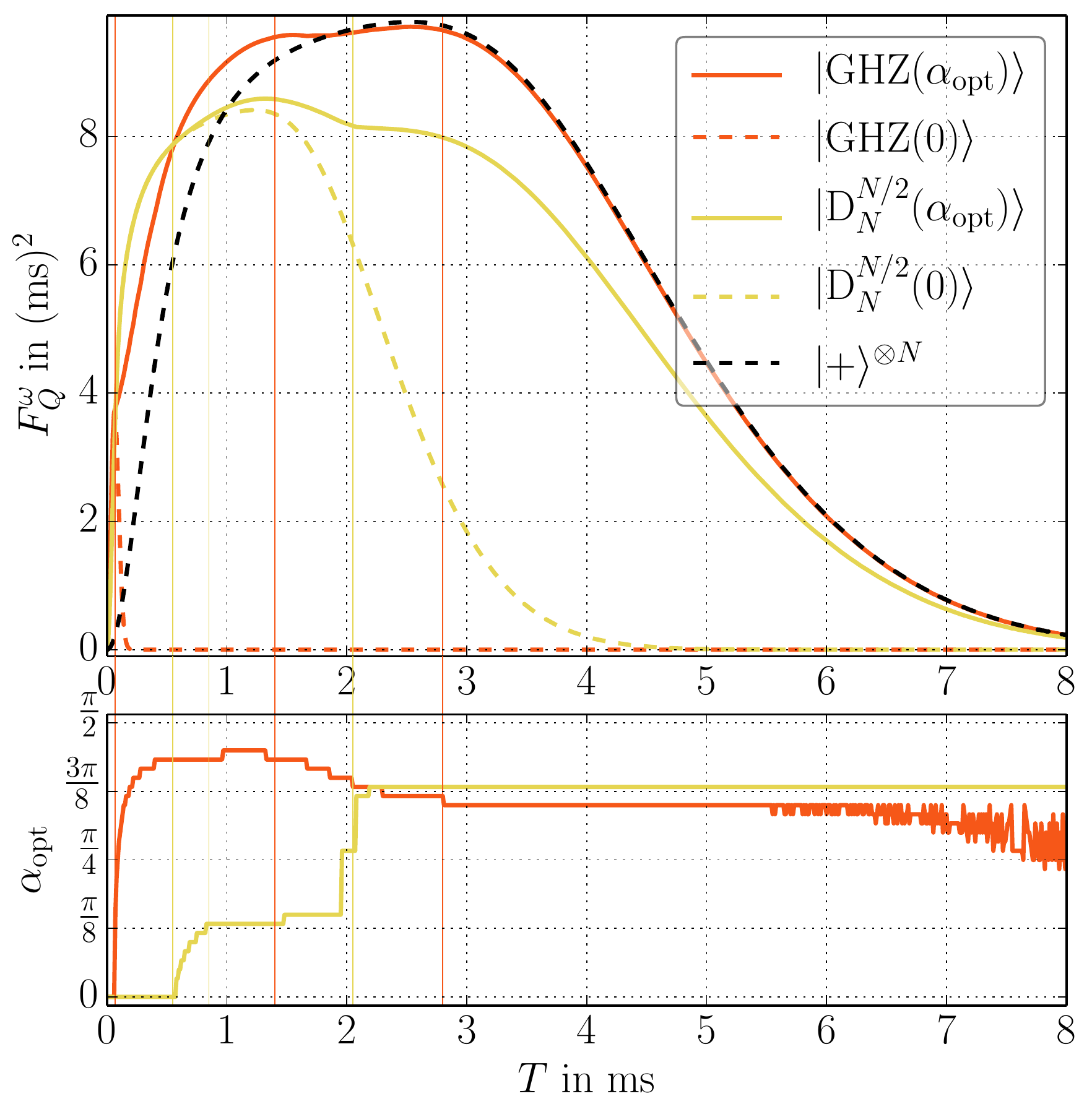}}
\caption{QFI for phase and frequency estimation with $N=50$ qubits.  The solid lines are the QFI optimized over the rotation angle $\alpha$ and dashed lines are the QFI of the original states.
\textbf{(a):} QFI for phase estimation over the time $T$ for different states.
  \textbf{(b):}  The upper plot shows the  QFI for frequency $\omega$ estimation. The lower plot shows the optimal rotation angle $\alpha_\mathrm{opt}$ over the time for the tested states. 
}\label{fig:Varianceforfreq_large_N}
\end{figure*}

{In this section we show that also for larger $N$ an optimization of the input states over the rotation angle $\alpha$ could lead to a higher precision. In Fig.~\ref{fig:Varianceforfreq_large_N} the QFI for phase- and frequency estimation with $N=50$ qubits by using the not rotated (dashed lines) and optimal rotated (solid lines) probe states is shown. The QFI for phase estimation decreases for all probe states faster than 
in Fig.~\ref{fig:Varianceforfreq}. Also the optimal rotation angle $\alpha_\mathrm{opt}$ changes in 
a smaller time scale at the beginning. For frequency estimation we find that the maximal QFI by 
using product states does not substantially change when comparing the estimation with $N=8$ 
and $N=50$ qubits. However, the QFI by using the optimal rotated GHZ state approaches the QFI by using a product state faster than in Fig.~\ref{fig:Varianceforfreq}.}

%%%%%%%%%%%%%%%%%%%%%%%%%%%%%%%%%%%%%%%%%%%%%%%%%%%%%%%%%%%%%%%%%
\section{Optimal states for DI in the steady state regime}\label{app:DI_optimal_state}
In the noiseless case, the maximal QFI is given by \cite{Giovannetti2006}
\begin{equation}
F_Q=4 (\lambda_{\mathrm{max}}-\lambda_{\mathrm{min}})^2,\label{eq:max_qfi}
\end{equation}
with $\lambda_{\mathrm{max}}$ ($\lambda_{\mathrm{min}}$) being the maximal (minimal) eigenvalue of the generator, here $\eins_{N_1} \otimes S_z^{N-N_1}$.
This maximal QFI can be reached with the state $ \ket{\Psi}=\left(\ket{v_{\mathrm{max}}}+\ket{v_{\mathrm{min}}}\right)/\sqrt{2}$, where $\ket{v_{\mathrm{max}}}$ and $\ket{v_{\mathrm{min}}}$ are eigenvectors of the generator $\eins_{N_1} \otimes S_z^{N-N_1}$ corresponding to the maximal and respectively minimal eigenvalue.
However, in presence of noise the initial state $\rho_0$ evolves, due to collective phase noise, into a mixed state until it becomes a mixture of states from the decoherence free subspace (DFS).
This mixed state does not change due to collective phase noise and is called steady state.
Now, we want to optimize the QFI in the steady state regime. Since the QFI  is convex that is
\begin{equation}
F_Q\left[p \varrho_1+(1-p)\varrho_2\right] \ge p F_Q \left[\varrho_1\right]+(1-p)F_Q \left[\varrho_2\right], 
\end{equation}
the QFI is maximal for pure states. 
Therfore, we have to maximize Eq. \eqref{eq:max_qfi} over all pure states $\ket{\Psi}$ lying in the DFS. In the DFS $\ket{v_{\max}}$ and $\ket{v_{\min}}$ need to have the same total number of excitations $k$ \cite{Lidar1998} and are given by
 \begin{align}
\ket{v_{\min}}= \begin{cases}
  \ket{1^{\otimes N_1}}\otimes\ket{1^{\otimes N_1-k} 0^{\otimes N-k}} &\mathrm{for}\,\,\, k>N_1,\\
  \ket{1^{\otimes k}0^{\otimes N_1-k}}\otimes\ket{ 0^{\otimes N-N_1}} &\mathrm{for}\,\,\,k\le N_1\\
 \end{cases}
 \end{align}
 and
 \begin{align}
\ket{v_{\max}}= \begin{cases}
  \ket{0^{\otimes N-k}1^{\otimes k-(N-N_1)} }\otimes\ket{1^{\otimes N-N_1}} &\mathrm{for}\,\,\, k>N-N_1,\\
  \ket{0^{\otimes N_1}}\otimes\ket{0^{\otimes N-N_1-k} 1^{\otimes k}} &\mathrm{for}\,\,\,k\le N-N_1.\\
 \end{cases}
 \end{align}
With these states, the QFI is given by
 \begin{align}
F_Q= \begin{cases}
 k^2 &\mathrm{for}\,\,\, k\le \mathrm{min}\{N_1,N-N_1\},\\
  N_1^2 &\mathrm{for}\,\,\,N_1<k\le N-N_1,\\
   (N-N_1)^2 &\mathrm{for}\,\,\,N-N_1 <k\le N_1,\\
    (N-k)^2 &\mathrm{for}\,\,\,k>\mathrm{max}\{N_1,N-N_1\},\\
 \end{cases}
 \end{align}
 which is maximal $F_Q=N^2/4 $ for $k=N_1=N-N_1=N/2$ and 
 the optimal state from the DFS is given by
 \begin{equation}
\ket{\Psi_{\mathrm{opt}}}=\frac{1}{\sqrt{2}}\left(\ket{\underbrace{0  \ldots 0}_{N/2}\underbrace{1  \ldots 1}_{N/2}}+\ket{\underbrace{1  \ldots 1}_{N/2}\underbrace{0  \ldots 0}_{N/2}}\right).
\end{equation}

%%%%%%%%%%%%%%%%%%%%%%%%%%%%%%%%%%%%%%%%%%%%%%%
\section{Scaling behaviour in the noiseless case by using DI}\label{app:DI_scaling_noiseless}
{
For phase estimation by using the metrological scheme from Fig.~\ref{fig:Metrology} (b), 
 the QFI for a pure initial state $\ket{\Psi}$ can be calculated analytically for the noiseless case by using the fact the QFI is additive under tensoring
\begin{equation}
F_Q\left[\varrho^{(1)}\otimes \varrho^{(2)}, A^{(1)}\otimes \eins+\eins \otimes A^{(2)}\right]=F_Q^\varphi\left[\varrho^{(1)}, A^{(1)}\right]+F_Q\left[ \varrho^{(2)}, A^{(2)}\right].
\end{equation}
For a product state $\ket{\Psi}=\ket{+}^{\otimes N/2} \otimes \ket{+}^{\otimes N/2}$ with $\ket{+}=(\ket{0}+\ket{1})/\sqrt{2}$  we find 
\begin{equation}
F_Q^\varphi=\frac{N}{2}.
\end{equation}
For a GHZ state $\ket{\Psi}=\ket{\mathrm{GHZ}} \otimes \ket{\mathrm{GHZ}}$ we find 
\begin{equation}
F_Q^\varphi=\left(\frac{N}{2}\right)^2.
\end{equation}
For the rotated  BSD state  $\ket{\mathrm{D}_{N/2}^{N/4}}_y \otimes \ket{\mathrm{D}_{N/2}^{N/4}}_y= U_x(\pi/2)\ket{\mathrm{D}_{N/2}^{N/4}} \otimes \ket{\mathrm{D}_{N/2}^{N/4}}$  the QFI is given by
\begin{equation}
F_Q^\varphi=\frac{N(N+4)}{8}.
\end{equation}}
%%%%%%%%%%%%%%%%%%%%%%%%%%%%%%%%%%%%%%%%%%%%%%%%%%%%%%%%%%

%%%%%%%%%%%%%%%%%%%%%%%%%%%%%%%%%%%%%%%%%%%%%%%%%%%%
\section{Scaling behaviour after dephasing by using DI}\label{app:DI_scaling_ss}
In Fig.~\ref{fig:phaseDI} we see that the QFI decreases with time to a constant greater than zero by using DI. In this section, we calculate this constant and investigate its scaling behaviour for the probe states.
The initial probe states evolve, due to collective phase noise into a mixed state. The steady state is a mixture in the decoherence free subspace and has therefore still some coherences. 
\subsection{Product state}\label{app:DI_scaling_ss_product}
The steady state for the product state $\ket{+}^{\otimes N}$ as an initial state is a mixture of symmetric Dicke states.
The non-zero eigenvalues are given by $\lambda_{k'}=\frac{C_N^{k'}}{2^N}$, where $C_N^{k'}=\tbinom{N}{k'}$ are binomial coefficients. The corresponding eigenvectors are given by $\ket{\mathrm{D}_N^{k'}}$. We can rewrite the symmetric Dicke states as
\begin{equation}
\ket{\mathrm{D}_N^{k'}}=\frac{1}{\sqrt{C_N^{k'}}}\sum_{q=0}^{k'} \sqrt{C_{N/2}^q}\ket{\mathrm{D}_{N/2}^q} \otimes \sqrt{C_{N/2}^{{k'}-q}}\ket{\mathrm{D}_{N/2}^{{k'}-q}}.
\end{equation}
With that formulation one finds that
\begin{align}
\begin{split}
\braket{\mathrm{D}_N^s|\eins \otimes S_z|\mathrm{D}_N^t}&= \frac{1}{\sqrt{C_N^sC_N^t}} \sum_{q,q'} \sqrt{C_{N/2}^q C_{N/2}^{q'} C_{N/2}^{s-q} C_{N/2}^{t-q'}} (t-q'-N/4)\\
&\cdot \braket{\mathrm{D}_{N/2}^q|\mathrm{D}_{N/2}^{q'}} \braket{\mathrm{D}_{N/2}^{s-q}|\mathrm{D}_{N/2}^{t-q'}}
\\
&= \frac{1}{\sqrt{C_N^sC_N^t}} \sum_{q} C_{N/2}^q\sqrt{ C_{N/2}^{s-q} C_{N/2}^{t-q}} (t-q-N/4)\braket{\mathrm{D}_{N/2}^{s-q}|\mathrm{D}_{N/2}^{t-q}}\\
&=  \frac{\delta_{s,t}}{C_N^s} \sum_{q} C_{N/2}^q C_{N/2}^{s-q} (s-q-N/4).\label{eq:dicke-dicke}
\end{split}
\end{align}

Based on these, we can rewrite the QFI by 
\begin{align}
F_Q= 4 \sum_{k'=0}^N \lambda_{k'} \sum_{k} \braket{\mathrm{D}_N^{k'}|\eins_{N/2}\otimes S_z^{N/2}|k}\braket{k|\eins_{N/2}\otimes S_z^{N/2}|\mathrm{D}_N^{k'}},
\end{align}
with $\ket{k}$ being eigenstates of $\varrho$ with $\lambda_{k}=0$ and  $\braket{k|\mathrm{D}_N^l}=0$ for all $l$. We can replace $\sum_{k} \ketbra{k}=\eins-\sum_l \ketbra{\mathrm{D}_N^l}$, such that the QFI reduces to
\begin{align}
\begin{split}
F_Q&= 4 \sum_{k'=0}^N \lambda_{k'}  \braket{\mathrm{D}_N^{k'}|\eins_{N/2}\otimes S_z^{N/2}|\left(\eins-\sum_l \ketbra{\mathrm{D}_N^l} \right)|\eins_{N/2}\otimes S_z^{N/2}|\mathrm{D}_N^{k'}}\\
&= 4 \sum_{k'=0}^N \lambda_{k'}  \left[\braket{\mathrm{D}_N^{k'}|\left(\eins_{N/2}\otimes S_z^{N/2}\right)^2|\mathrm{D}_N^{k'}}- \left(\sum_l \braket{\mathrm{D}_N^l|\eins_{N/2}\otimes S_z^{N/2}|\mathrm{D}_N^{k'}}\right)^2\right].
\end{split}
\end{align}
And with Eq. \eqref{eq:dicke-dicke} and because of the symmetry of the state we can express the expectation value
\begin{equation}
\braket{\mathrm{D}_N^k|\eins_{N/2}\otimes S_z^{N/2}|\mathrm{D}_N^k}=\braket{\mathrm{D}_N^k|S_z^{N/2}\otimes \eins_{N/2}|\mathrm{D}_N^k}=\frac{1}{2}\braket{\mathrm{D}_N^k| S_z^{N}|\mathrm{D}_N^k}= \frac{k-N/2}{2}.
\end{equation} 
Replacing the second term leads to
\begin{align}
\begin{split}
F_Q&= 4 \sum_{k'=0}^N \lambda_{k'} \left[ \braket{\mathrm{D}_N^{k'}|\left(\eins_{N/2}\otimes S_z^{N/2}\right)^2|\mathrm{D}_N^{k'}}- \left(\frac{k'-N/2}{2}\right)^2\right]\\
&= \frac{1}{2^{N-2}}\sum_{k'=0}^N \left[ \sum_{q=0}^{k'} C_{N/2}^q C_{N/2}^{k'-q} \left(k'-q-\frac{N}{4}\right)^2- C_N^{k'}\left(\frac{k'-N/2}{2}\right)^2\right]\\
&=\frac{1}{2^{N-2}}\sum_{k'=0}^N \left[ \sum_{q=0}^{k'} C_{N/2}^q C_{N/2}^{k'-q} \left(k'-q-\frac{N}{4}\right)^2- \sum_{q=0}^{k'} C_{N/2}^q C_{N/2}^{k'-q}\left(\frac{k'-N/2}{2}\right)^2\right]\\
&=\frac{1}{2^{N-2}} \sum_{k'=0}^N  \sum_{q=0}^{k'} C_{N/2}^q C_{N/2}^{k'-q} \frac{\left(k'-2q\right)\left(3k'-N-2q\right)}{4} \\
&=\frac{1}{2^{N-2}} N 2^{N-4}=N/4.
\end{split}
\end{align}

\subsection{GHZ state}\label{app:DI_scaling_ss_GHZ}
The steady state for a GHZ state as an initial state is given by
\begin{align}
\begin{split}
(\vr_{GHZ}\otimes\vr_{GHZ})_{\mathrm{steady state}}&= \frac{1}{4}(\ketbra{\underbrace{0  \ldots 0}_{N}}+\ketbra{\underbrace{1  \ldots 1}_{N}}\\
&+\ketbra{\underbrace{0  \ldots 0}_{N/2} \underbrace{1  \ldots 1}_{N/2}}
+\ketbra{\underbrace{1  \ldots 1}_{N/2}\underbrace{0  \ldots 0}_{N/2} }\\
&+\KetBraO{\underbrace{1  \ldots 1}_{N/2}\underbrace{0  \ldots 0}_{N/2}}{\underbrace{0  \ldots 0}_{N/2}\underbrace{1  \ldots 1}_{N/2}}{}+\KetBraO{\underbrace{0  \ldots 0}_{N/2}\underbrace{1  \ldots 1}_{N/2}}{\underbrace{1  \ldots 1}_{N/2}\underbrace{0  \ldots 0}_{N/2}}{}
).
\end{split}
\end{align}
There are four remarkable eigenvectors; 
\begin{align}
\begin{split}
&\ket{v_1}=\ket{\underbrace{0  \ldots 0}_{N}},\\
&\ket{v_2}=\ket{\underbrace{1  \ldots 1}_{N}},\\
&\ket{v_3}=1/\sqrt{2} (\ket{\underbrace{0  \ldots 0}_{N/2} \underbrace{1  \ldots 1}_{N/2}}+\ket{\underbrace{1  \ldots 1}_{N/2}\underbrace{0  \ldots 0}_{N/2} }),\\
&\ket{v_4}=1/\sqrt{2} (-\ket{\underbrace{0  \ldots 0}_{N/2} \underbrace{1  \ldots 1}_{N/2}}+\ket{\underbrace{1  \ldots 1}_{N/2}\underbrace{0  \ldots 0}_{N/2} })
\end{split}
\end{align}
with eigenvalues $\lambda_{1}=\lambda_{2}=1/4$, $\lambda_{3}=1/2$ and $\lambda_{4}=0$. All other eigenvalues $\lambda_{5\ldots 2^N}=0$ and we denote the eigenvectors corrosponding to these eigenvalues with $\ket{v_{5\ldots 2^N}}$.
It is easy to show, that 
\begin{align}
\begin{split}
&\braket{v_1|\eins_{N/2}\otimes S_z^{N/2}|v_3}=\braket{v_1|\eins_{N/2}\otimes S_z^{N/2}|v_4}=0,\\
&\braket{v_2|\eins_{N/2}\otimes S_z^{N/2}|v_3}=\braket{v_2|\eins_{N/2}\otimes S_z^{N/2}|v_4}=0,\\
&\braket{v_{1\ldots 4 }|\eins_{N/2}\otimes S_z^{N/2}|v_{5\ldots 2^N}}=0
\end{split}
\end{align}
 and all terms with the same eigenvalues also vanish, so that the sum in the QFI reduces to
\begin{align}
F_Q= 4\cdot \frac{(\lambda_3 -\lambda_4)^2}{\lambda_3 +\lambda_4} \left|\braket{v_3|\eins_{N/2}\otimes S_z^{N/2}|v_4}\right|^2= \frac{N^2}{8}.
\end{align}
with $\eins_{N/2}\otimes S_z^{N/2}\ket{v_3}=N/4 \ket{v_4}$.

\subsection{Bipartite symmetric Dicke state in the x-basis}\label{app:DI_scaling_ss_Dicke}
We start with an arbitrary symmetric Dicke state in the $x$ basis for both inputs. We can express this BSD state in the basis of symmetric Dicke states in the $z$ basis by
\begin{align}
\begin{split}
\ket{\mathrm{D}_{N_1}^{k_1},\mathrm{D}_{N-N_1}^{k_2}}_x=\sum_{k_1',k_2'} \braket{\mathrm{D}_{N_1}^{k_1'}|U_y^{N_1}\left(\frac{\pi}{2}\right)|\mathrm{D}_{N_1}^{k_1}}\braket{\mathrm{D}_{N-N_1}^{k_2'}|U_y^{N-N_1}\left(\frac{\pi}{2}\right)|\mathrm{D}_{N-N_1}^{k_2}}\ket{\mathrm{D}_{N_1}^{k_1'},\mathrm{D}_{N-N_1}^{k_2'}}.
\end{split}
\end{align}
 For simplicity we choose here a rotation around the $y$-axis.
Where $d^N_{k',k}\left(\frac{\pi}{2}\right)=\braket{\mathrm{D}_{N}^{k'}|U_y^{N}\left(\frac{\pi}{2}\right)|\mathrm{D}_{N}^{k}}:=d^N_{k',k}$ is the ''small'' Wigner $D$ matrix \cite{Wigner1932} for a rotation angle of $\pi/2$

\begin{align}
d^N_{k',k}=\sqrt{\frac{C_N^{k}}{C_N^{k'} 2^N}}\sum_{s=\max\{0,k'-k\}}^{\min\{N-k,k'\}} (-1)^{k'-k+s}\frac{ C_{N-k}^{s}}{C_{k}^{k'-s}}.
\end{align}
Now we can rewrite the state as
\begin{align}
\ket{\mathrm{D}_{N_1}^{k_1},\mathrm{D}_{N-N_1}^{k_2}}_x=\sum_{k_1',k_2'}d^{N_1}_{k_1',k_1}d^{N-N_1}_{k_2',k_2}\ket{\mathrm{D}_{N_1}^{k_1'},\mathrm{D}_{N-N_1}^{k_2'}}.
\end{align}
For a fixed number of excitations $k'=k_1'+k_2'$ we have
\begin{align}
\begin{split}
\ket{\mathrm{D}_{N_1}^{k_1},\mathrm{D}_{N-N_1}^{k_2}}_x&=\sum_{k'=0}^{N} \sum_{q=\max\{k'-N_1,0\}}^{\min\{k',N_1\}}d^{N_1}_{q,k_1}d^{N-N_1}_{k'-q,k_2}\ket{\mathrm{D}_{N_1}^{q},\mathrm{D}_{N-N_1}^{k'-q}}\\
&=\sum_{k'=0}^{N} \ket{l_{k'}}=\sum_{k'=0}^{N} \sqrt{p_{k'}}\ket{v_{k'}},
\end{split}
\end{align}
with the not normalized states $\ket{l_{k'}}$ and the normalized states $\braket{v_{k'}|v_{k'}}=1$.
We can calculate the probability $p_{k'}$ for being in the state $\ket{v_{k'}}$ by 
\begin{align}
p_{k'}=\braket{l_{k'}|l_{k'}}=\sum_{q=\max\{k'-N_1,0\}}^{\min\{k',N_1\}}\left(d^{N_1}_{q,k_1}d^{N-N_1}_{k'-q,k_2}\right)^2.
\end{align}
With those we find the normalized states
\begin{align}
\ket{v_{k'}}=\frac{1}{\sqrt{p_{k'}}}\sum_{q=\max\{k'-N_1,0\}}^{\min\{k',N_1\}}d^{N_1}_{q,k_1}d^{N-N_1}_{k'-q,k_2}\ket{\mathrm{D}_{N_1}^{q},\mathrm{D}_{N-N_1}^{k'-q}}.
\end{align}
With the probabilities $p_{k'}$ and the states $\ket{v_{k'}}$, we can write the rotated state in the $z$ basis as
\begin{align}
\varrho=\sum_{m,n} \sqrt{p_m p_n} \ket{v_m}\bra{v_n}.
\end{align}
After dephasing, only the elements with $m=n$ remain \cite{Lidar1998} such that the steady state is given by $\varrho_{f}=\sum_m p_m \ketbra{v_m}$. The non-zero eigenvalues of this state are $\lambda_{k'}=p_{k'}$ with the corresponding eigenvectors $\ket{v_{k'}}$.
Now we can show, that 
\begin{align}
\begin{split}
\braket{v_s|\eins_{N_1}\otimes S^{N-N_1}_z|v_{k'}}&=
\frac{1}{\sqrt{p_{k'} p_s}}\sum_{q=\max\{k'-N_1,0\}}^{\min\{k',N_1\}}\sum_{q'=\max\{N-N_1-s,0\}}^{\min\{s,N_1\}}\!\!\!\!\!\!\!\!\!\!\!\!\!\!\!\!d^{N_1}_{q,k_1}d^{N-N_1}_{k'-q,k_2}d^{N_1}_{q',k_1}d^{N-N_1}_{s-q',k_2}\underbrace{\braket{\mathrm{D}_{N_1}^{q'}|\mathrm{D}_{N_1}^{q}}}_{\delta_{q,q'}}\!\!\!\!\underbrace{\braket{\mathrm{D}_{N-N_1}^{s-q'}|S^{N-N_1}_z|\mathrm{D}_{N-N_1}^{k'-q}}}_{(k'-q-N-N_1/2)\braket{\mathrm{D}_{N-N_1}^{s-q'}|\mathrm{D}_{N-N_1}^{k'-q}}}\\
&\propto \delta_{s,k'}.
\end{split}
\end{align}
Such that the QFI reduces to
\begin{align}
F_Q&=4 \sum_{k'=0}^N \lambda_{k'} \sum_{k} \braket{v_{k'}|\eins_{N_1}\otimes S_z^{N-N_1}|k}\braket{k|\eins_{N_1}\otimes S_z^{N-N_1}|v_{k'}}
\end{align}
with $\ket{k} \neq \ket{v_{k}}$ being an eigenvector with a zero eigenvalue. Now we repeat the same steps as for product states as initial states to rewrite the QFI as
\begin{align}
F_Q&=4 \sum_{k'=0}^N \lambda_{k'} \left(\Delta_{v_{k'}} (\eins_{N_1} \otimes S_z^{N-N_1})\right)^2,
\end{align}
where $\left(\Delta_{v_{k'}} (\eins_{N_1} \otimes S^{N-N_1}_z)\right)^2$ denotes the variance and is given by
\begin{align}
\begin{split}
\left(\Delta_{v_{k'}} (\eins_{N_1} \otimes S_z^{N-N_1})\right)^2&=
\frac{1}{\lambda_{k'}} \sum_{q=\max\{k'-N_1,0\}}^{\min\{k',N_1\}}\left(d^{N_1}_{q,k_1}d^{N-N_1}_{k'-q,k_2}\right)^2\left(k'-q-\frac{N-N_1}{2}\right)^2 \\
&-\frac{1}{\lambda_{k'}^2} \left[\sum_{q=\max\{N-N_1-k',0\}}^{\min\{k',N_1\}}\left(d^{N_1}_{q,k_1}d^{N-N_1}_{k'-q,k_2}\right)^2 \left(k'-q-\frac{N-N_1}{2}\right)\right]^2.\end{split}\label{eq:general_Dicke_var}
\end{align}

Together the QFI is given by
\begin{align}
\begin{split}
F_Q^\varphi[\vr_\mathrm{f}] &= 4 \sum_{k'=0}^{N} \left\lbrace\sum_{q=\max\{N-N_1-k',0\}}^{\min\{k',N_1\}}\left(d^{N_1}_{q,k_1}d^{N-N_1}_{k'-q,k_2}\right)^2\left(k'-q-\frac{N-N_1}{2}\right)^2 \right.  \\
&\left. -\frac{\left[\sum_{q=\max\{N-N_1-k',0\}}^{\min\{k',N_1\}}\left(d^{N_1}_{q,k_1}d^{N-N_1}_{k'-q,k_2}\right)^2 \left(k'-q-\frac{N-N_1}{2}\right)\right]^2}{\sum_{q=\max\{N-N_1-k',0\}}^{\min\{k',N_1\}}\left(d^{N_1}_{q,k_1}d^{N-N_1}_{k'-q,k_2}\right)^2 }  \right\rbrace ,
\end{split}\label{eq:QFI_for_Dicke}
\end{align}
This is a general formula for the QFI after dephasing for an initial state of the form $\ket{\mathrm{D}_{N_1}^{k_1},\mathrm{D}_{N-N_1}^{k_2}}_x$.
From Fig. \ref{fig:max_FI_Dicke}, we see, that Eq. \eqref{eq:QFI_for_Dicke} is maximal for the probe state with $N_1=N-N_1=N/2$ and $k_1=k_2=N/4$, where $N=4j$, with $j$ being an integer.
For this simple case and $N \le 1000$ we have verified that the formula in Eq. \eqref{eq:QFI_for_Dicke} is equivalent to
 \begin{equation}
 F_Q=\frac{N(N+4)}{16}.
 \end{equation}
 It is very likely to hold also in general, but has not been proven yet.

 \section{Optimization for product states}\label{app:DI_optimization_product}
 We want to investigate Eq. \eqref{eq:QFI_for_Dicke} for the case of $k_1=k_2=0$. This means, that the input probe state is a product state. For this case we optimize the splitting $N_1$ and $N-N_1$.

 For $k_1=k_2=0$ we find that $\left(d^{N_1}_{q,0}\right)^2=C_{N_1}^{q} 2^{-N_1}$ such that 
 \begin{align}
 \left(d^{N_1}_{q,0}d^{N-N_1}_{k'-q,0}\right)^2=C_{N_1}^{q}C_{N-N_1}^{k'-q} 2^{-N}.
 \end{align}
 Then the eigenvalues are given by
 \begin{align}
 \lambda_{k'}=2^{-N}\sum_{q=\max\{k'-N_1,0\}}^{\min\{k',N_1\}}C_{N_1}^{q}C_{N-N_1}^{k'-q}.
 \end{align}
 We can split the sum for $k'\le N_1$ and $k'\ge N_1$,
  \begin{align}
 \lambda_{k'}= \begin{cases}
 2^{-N}\sum_{q=0}^{k'}C_{N_1}^{q}C_{N-N_1}^{k'-q} &\mathrm{for}\,\,\, k'\le N_1,\\
  2^{-N}\sum_{q=k'-N_1}^{N_1}C_{N_1}^{q}C_{N-N_1}^{k'-q} &\mathrm{for}\,\,\,k'\ge N_1.\\
 \end{cases}
 \end{align}
 This expression can be simplified by using $\sum_{q=0}^{k'}C_{N_1}^{q}C_{N-N_1}^{k'-q}=C_N^{k'}$ and shifting the summation $q=j+(k'-N_1)$ for the case $k'\ge N_1$ such that $\sum_{q=k'-N_1}^{N_1}C_{N_1}^{q}C_{N-N_1}^{k'-q}=\sum_{j=0}^{N-k'}C_{N_1}^{(N-k')-j}C_{N-N_1}^{j}=C_{N}^{N-k'}=C_N^{k'}$. For both cases the eigenvalues are given by
 \begin{align}
 \lambda_{k'}=2^{-N}C_N^{k'}.
 \end{align}
 Next, we can simplify the second term in Eq. \eqref{eq:general_Dicke_var} by
 \begin{align}
 \begin{split}
  \braket{\eins \otimes S_z}^2&=\frac{1}{\lambda_{k'}^2} \left[\sum_{q=\max\{N-N_1-k',0\}}^{\min\{k',N_1\}}\left(d^{N_1}_{q,k_1}d^{N-N_1}_{k'-q,k_2}\right)^2 \left(k'-q-\frac{N-N_1}{2}\right)\right]^2\\
  &=\frac{1}{\lambda_{k'}^2} \left[\sum_{q=\max\{N-N_1-k',0\}}^{\min\{k',N_1\}}2^{-N}C_{N_1}^{q}C_{N-N_1}^{k'-q} \left(k'-q-\frac{N-N_1}{2}\right)\right]^2.
  \end{split}
 \end{align}
We again split the sum in two cases $k'\le N_1$ and $k'\ge N_1$. For $k'\le N_1$ we find
 \begin{align}
 \begin{split}
  \braket{S_z}^2
  &=\frac{1}{\lambda_{k'}^2} \left[\sum_{q=0}^{k'}2^{-N}C_{N_1}^{q}C_{N-N_1}^{k'-q} \left(k'-q-\frac{N-N_1}{2}\right)\right]^2\\
  &=\left[\frac{(-1)^{1+k'}(2k'-N)(N-N_1)(-1+k'-N)!}{2 C_N^{k'} (k')! (-N)!}\right]^2
  =\left[\frac{(2k'-N)(N-N_1)}{2 N}\right]^2,
  \end{split}
 \end{align}
 with $(-N)!=\Gamma(-N+1)$.
For the case $k'\ge N_1$, shifting the summation with $q=j+(k'-N_1)$ like for the eigenvalues and simplifying in the same way leads to the same result.
Now we can simplify the expression for the QFI in Eq. \eqref{eq:QFI_for_Dicke} by
\begin{align}
\begin{split}
F_Q &= 4 \sum_{k'=0}^{N} \left\lbrace \sum_{q=\max\{N-N_1-k',0\}}^{\min\{k',N_1\}}\left(d^{N_1}_{q,k_1}d^{N-N_1}_{k'-q,k_2}\right)^2\left(k'-q-\frac{N-N_1}{2}\right)^2 \right. 
\\
& \left. -\frac{1}{\lambda_{k'}} \left[\sum_{q=\max\{N-N_1-k',0\}}^{\min\{k',N_1\}}\left(d^{N_1}_{q,k_1}d^{N-N_1}_{k'-q,k_2}\right)^2 \left(k'-q-\frac{N-N_1}{2}\right)\right]^2\right\rbrace
\\
&= 4 \sum_{k'=0}^{N}\sum_{q=\max\{N-N_1-k',0\}}^{\min\{k',N_1\}}2^{-N} C_{N_1}^{q}C_{N-N_1}^{k'-q}\left(k'-q-\frac{N-N_1}{2}\right)^2-2^{-N}C_N^{k'} \left[\frac{(2k'-N)(N-N_1)}{2 N}\right]^2\\
&=4 \sum_{k'=0}^{N}\sum_{q=\max\{N-N_1-k',0\}}^{\min\{k',N_1\}}2^{-N} C_{N_1}^{q}C_{N-N_1}^{k'-q}\left[\left(k'-q-\frac{N-N_1}{2}\right)^2-\left(\frac{(2k'-N)(N-N_1)}{2 N}\right)^2\right]\\
&=4 \sum_{k'=0}^{N}\sum_{q=\max\{N-N_1-k',0\}}^{\min\{k',N_1\}}2^{-N} C_{N_1}^{q}C_{N-N_1}^{k'-q}\left\lbrace\frac{(N q-k' N_1)\left[k' (N_1-2N)+N(N-N_1+q)\right]}{N^2}\right\rbrace.
\end{split}
\end{align}
Again, we split the summation over $q$ into two cases $k'\le N_1$ and $k'\ge N_1$. For $k'\le N_1$ we find
\begin{align}
\begin{split}
& \sum_{q=\max\{N-N_1-k',0\}}^{\min\{k',N_1\}}2^{-N} C_{N_1}^{q}C_{N-N_1}^{k'-q}\left\lbrace\frac{(N q-k' N_1)\left[k' (N_1-2N)+N(N-N_1+q)\right]}{N^2}\right\rbrace\\
&=2^{-N} C_{N}^{k'} \frac{k'(N-k')(N-N_1)N_1}{(N-1)N^2}=\lambda_{k'}\frac{k'(N-k')(N-N_1)N_1}{(N-1)N^2}.
\end{split}
\end{align}
For the case $k'\ge N_1$, shifting the summation with $q=j+(k'-N_1)$ like for the eigenvalues and simplifying in the same way leads to the same result, such that we can calculate the variance to
\begin{equation}
\left(\Delta_{v_{k'}} \eins \otimes S_z\right)^2=\frac{k'(N-k')(N-N_1)N_1}{(N-1)N^2}.
\end{equation}
Together, the QFI is given by
\begin{align}
\begin{split}
F_Q &= 4 \sum_{k'=0}^{N}\lambda_{k'}\left(\Delta_{v_{k'}} \eins_{N_1} \otimes S^{N-N_1}_z\right)^2\\
&=4 \sum_{k'=0}^{N}2^{-N}C_N^{k'}\frac{k'(N-k')(N-N_1)N_1}{(N-1)N^2}= \frac{(N-N_1)N_1}{N},
\end{split}
\end{align}
which is maximal $F_Q^{\max}=N/4$ for $N_1=N/2$.

 \end{appendix}
 
 \twocolumngrid

\end{document}